# Significance of nuclear quantum effects in hydrogen bonded molecular chains

Aleš Cahlík[1,2,3†], Jack Hellerstedt[1,†], Jesús I. Mendieta-Moreno[1†], Martin Švec[1,3], Vijai M. Santhini[1], Simon Pascal[4], Diego Soler-Polo[5], Sigurdur I. Erlingsson[6], Karel Výborný[1], Pingo Mutombo[1,8], Ondrej Marsalek[7], Olivier Siri[4*], Pavel Jelínek[1,3*]

[1] Institute of Physics of the Czech Academy of Sciences, v.v.i., Cukrovarnicka 10, CZ-16200 Prague 6, Czech Republic

[2] Faculty of Nuclear Sciences and Physical Engineering, Czech Technical University in Prague, Břehová 78/7, CZ-11519 Prague 1, Czech Republic

[3] Regional Centre of Advanced Technologies and Materials, Palacký University, Šlechtitelů 27, CZ-78371 Olomouc, Czech Republic

[4] Aix Marseille Univ, CNRS, CINaM, UMR 7325, Campus de Luminy, F-13288 Marseille Cedex 09, France

[5] Universidad Autónoma de Madrid, Campus Cantoblanco, Madrid, Spain

[6] School of Science and Engineering, Reykjavik University, Menntavegi 1, IS-101 Reykjavik, Iceland

[7] Charles University, Faculty of Mathematics and Physics, Ke Karlovu 3, CZ-12116 Prague 2, Czech Republic

[8] Department of Petrochemistry and Refining, University of Kinshasa, Kinshasa, Democratic Republic of Congo

*Correspondence to: olivier.siri@univ-amu.fr, jelinekp@fzu.cz

†These authors contributed equally to this work.

**Abstract:**

In hydrogen bonded systems, nuclear quantum effects such as zero-point motion and tunneling can significantly affect their material properties through underlying physical and chemical processes. Presently, direct observation of the influence of nuclear quantum effects on the strength of hydrogen bonds with resulting structural and electronic implications remains elusive, leaving opportunities for deeper understanding to harness their fascinating properties.

We studied hydrogen-bonded one-dimensional quinonediimine molecular networks which may adopt two isomeric electronic configurations via proton transfer. Herein, we demonstrate that concerted proton transfer promotes a delocalization of π-electrons along the molecular chain, which enhances the cohesive energy between molecular units, increasing the mechanical stability of the chain and giving rise to new electronic in-gap states localized at the ends.

These findings demonstrate the identification of a new class of isomeric hydrogen bonded molecular systems where nuclear quantum effects play a dominant role in establishing their chemical and physical properties. We anticipate that this work will open new research directions towards the control of mechanical and electronic properties of low-dimensional molecular materials via concerted proton tunneling.

**TEXT:**

Nuclear quantum effects (NQEs), such as proton tunneling and zero-point motion can play an important role in understanding structural[1] and material properties[2–4] of hydrogen-bonded systems at low temperatures. It has been demonstrated both theoretically[5] and experimentally[6] that nuclear quantum effects may have a pronounced two-fold effect on the strength of hydrogen bonds, either further weakening of already weak hydrogen bonds or conversely, strengthening of the relatively strong ones. NQEs can induce strong proton delocalization with direct consequences on chemical activity of the system.[7] Complex, concerted many-body proton motion in ice has been described both experimentally[8] and theoretically[9,10]. In this context, recent progress in scanning probe microscopy providing unprecedented spatial resolution on single molecules via a proper tip functionalization[11,12] has enabled the direct observation of concerted proton motion in water tetramers[13].

Despite these advances, our present understanding of NQEs remains incomplete. In this work, we show that the concerted proton motion in a H-bonded 1D molecular system not only enhances its mechanical stability but directly modifies its electronic structure, forming new electronic in-gap states localized at the ends of the chain.

2,5-diamino-1,4-benzoquinonediimines (DABQDI, structure in Fig.1a inset) belong to a family of quinoid molecules with intriguing electronic properties[14] stemming from a unique distribution of their π-electrons. DABQDI quinones contain 12 π-electrons which can be perceived as two independent π-subsystems containing 6 conjugated π-electrons (the nitrogen lone pair is conjugated with the two double bonds), chemically linked via two C-C σ-bonds, but electronically not conjugated[15,16]. The molecular DABQDI building blocks exist in solution as two tautomers in equilibrium, whose mutual alternation can be realized via a fast intramolecular double proton transfer that generates a structure of higher symmetry (i.e. an averaged form of the two tautomers) which directly alters the π-conjugation of the whole DABQDI molecule.[17] Curiously, although unsubstituted DABQDI (N-H) was reported in the literature in 1887[18], this molecule has since rarely been investigated, probably due to its very low solubility and poor stability (co-condensation, hydrolysis, and oxidation side reactions).[19]

Here we explore self-assembled molecular chains built from the precursor DABQDI on a metallic Au(111) surface at low temperatures (5 K) under ultra-high vacuum conditions. The presence of imine (as H-acceptor) and amine groups (as H-donor) enables, in principle, the formation of 1D intermolecular hydrogen-bonded assemblies. Such chains may adopt two isomeric π-conjugations resulting from the distinct alternation of double and single bonds according to the position of the amine group hydrogen atoms. In principle, the energy landscape of the system can be mimicked by a symmetric double-well potential, which has different ground states in either the classical or quantum picture. While in the classical picture the system is localized in one of the wells, the quantum ground state exists as a superposition of two states[20]. As we will show later on, the quantum state strongly affects the electronic structure of the chain. In this one-dimensional configuration, the presence of concerted proton transfer not only induces resonant tunneling between the two degenerate π-conjugated electronic states of the chain, but it also mediates an effective coupling of the π-electron systems across the chain. This coupling of resonant electronic configurations leads to the emergence of new electronic states and reinforcement of the mechanical stability of the molecular chain.

Figure 1a shows a representative overview scanning tunneling microscopy (STM) image acquired at 5 K of linear self-assembled 1D molecular structures. The chains form upon deposition of single DABQDI molecules onto a Au(111) substrate held at low temperature (5 K, AFM image of single DABQDI shown in Fig. S1c) which is subsequently warmed to room temperature where the chains self-assemble via surface diffusion (for detailed description of the sample preparation see the Supplementary Information). Typically, we observe chains with lengths ranging between 3 and 100 nm, oriented independently of the surface herringbone reconstruction. The hallmark of these molecular chains is the presence of characteristic bright spots in STM images located at the ends of the chains, as can be seen in Fig. 1b. In addition to the chains, distinct individual molecular species are present on the substrate, predominantly situated on the herringbone elbows, which we identify as individual molecules containing an extra proton as will be discussed later.

We are readily able to contact and manipulate complete chains along the surface by approaching the tip to a chain end with a subsequent lateral tip movement. Figure 1c displays a series of STM images of the same chain acquired between consecutive manipulations (see also Movie S1). Chains always remained intact during manipulation without loss of their structural integrity while preserving the bright spots at their ends (Fig. S2). This demonstrates not only a weak dispersive interaction between the molecular chains and the underlying metallic surface, but more importantly a relatively strong intermolecular binding.

The picture can qualitatively change when the DABQDI molecules are sublimed onto the surface at room temperature and subsequently cooled down to 5 K for imaging. For certain preparation conditions the resulting molecular chains incorporate more defects, their growth is restricted by the herringbone reconstruction of the Au(111) surface and they lack bright end terminations, as can be seen in Fig. S3a, b, c. Moreover, the mechanical stability is drastically reduced, making lateral manipulation impossible. Instead, mechanical interaction with the scanning probe easily splits the chains into segments, as shown in Fig. S3d. One possibility to explain the difference between the mechanical properties of the two types of chain highlights is the impact of proton tunneling on the strength of hydrogen bonding between the molecular units forming these chains. Indeed, the fact that NQEs may further enhance cohesion of relatively strong hydrogen bonds has been explored theoretically[5,6], but direct observations are lacking. For clarity, in the rest of the manuscript we will refer to the exemplary former species as symmetric chains while the latter will be referred to as canted chains (n.b. experimental control of the formation of symmetric vs. canted chains is imperfect: see Methods and extended discussion in the Supplementary Information).

To better understand the internal structure of the adsorbed chains and single molecular species, we acquired high-resolution atomic force microscopy (AFM) images with a CO-functionalized probe[11,12]. This scanning probe technique has repeatedly demonstrated the unique capabilities of unambiguous discrimination of chemical [21, 22] and atomic[23] structure, electrostatic potential mapping,[24] or identification of the spin state[25] of single molecules on surfaces.

The high-resolution AFM image of a single molecular species, not participating in the chain formation, shown in Fig. 2a, reveals a characteristic trapezoidal shape. Perfect agreement of the experimental AFM image with the simulation based on the chemical structure shown in Fig. 2b can be achieved by including the presence of an extra hydrogen in the single molecule species (Fig. 2c, d). The detailed structure of the hydrogenated molecule as well as the origin of the extra hydrogen are discussed in the Supplementary Information. We presume this additional hydrogen impedes an efficient self-assembling process via hydrogen bonding with the remaining molecules (the H-acceptor capability is suppressed).

Figure 2e presents high resolution AFM images of the symmetric chain interior. By registering the molecular structure with the underlying Au(111) surface (see Fig. S4) we determined an incommensurable alignment of the molecular chains with the substrate and the distance between two contiguous molecules to be $8.0 \pm 0.1$ Å. This excludes the possibility of covalent bonding between nitrogen atoms, since the Density Functional Theory (DFT)[26–29] calculated periodicity of a chain composed of covalently bonded molecules is significantly lower (5.7 Å, see Fig. S5e). The fact that our experiments are carried out in UHV conditions significantly reduces the possibility of contamination. A natural explanation of the large mechanical stability of symmetric chains would be a formation of organometallic chains with gold adatoms. However, this scenario can be ruled out by several observations. Namely, in rare instances we observe the formation of short defective chains (see Fig. S5a), whose AFM contrast is clearly distinct from the straight chain since it shows a characteristic "x-like" feature in between the molecular units. Such contrast feature fits well to simulated AFM[30,31] images of a fully optimized metal-organic Au-DABQDI chain with four-fold coordinated gold atoms on a Au(111) surface obtained from total energy DFT calculations (Fig. S5b, c). More elaborate discussion ruling out the presence of the gold organometallic chains and the other possible scenario, the gain or loss of additional hydrogens, can be found in the Supplemental Material.

A classical total energy DFT calculation of a molecular DABQDI chain assembled by hydrogen bonds (Fig. 2k, l) provides relatively good agreement but with slightly larger distance between two adjacent molecules (8.12 Å), further disfavoring the dative hypothesis. However, molecules in such chains are canted with respect to the main chain axis in order to decrease the energy by aiming the hydrogens participating in the hydrogen bond toward their respective nitrogen atoms. This is in contrast with the experimentally observed symmetric chain structure where all the molecular units are symmetric around the main axis (see Fig. 2e and Fig. S6a). We resolved this inconsistency by using Path Integral Molecular Dynamics (PIMD) simulations (details later) which account for NQEs (corresponding atomic structure shown in Fig. 2g). The calculated intermolecular distance using average atomic positions of PIMD calculations, 8.03 Å, fits very well to the experimental value (8.0 ± 0.1 Å). Moreover, the corresponding simulated AFM image (Fig. 2f) shows a highly symmetric arrangement caused by slight rearrangement of the positions of hydrogen and nitrogen atoms driven by the proton tunneling, which agrees with the experimental evidence.

On the other hand, the high-resolution AFM images acquired on canted chains (see Fig. 2i and Fig. S6b) match the AFM simulation (Fig. 2j) of the canted structure predicted by the total energy DFT simulations (Fig. 2k). This indicates that in the canted chains the molecular units are frozen in one of the two possible configurations with lower energy due to an external constraint, while the symmetric chains are a superposition of the two degenerate electronic states that is driven by proton tunneling between adjacent nitrogen atoms. One possible way to confirm the relevance of the proton tunneling would be to carry out the same experiment with molecules synthesized using six deuterium atoms. Unfortunately, performing the same experiment with deuterated DABQDI precursor proved to be unfeasible, mainly due to its poor stability (discussed in the Supplemental Material).

To overcome this experimental limitation, we have performed PIMD[32,33] simulations in order to elucidate the importance of NQEs and their impact on the structural properties of the molecular chains. We analyzed the results obtained from QM/MM (Quantum Mechanics/Molecular Mechanics) simulations[34–38] at different temperatures in which the quinone molecules were included in the QM region and the metallic surface is in the classical region. Figure 3a shows a free energy profile[39] using our QM/MM (DFT) method treating all the nuclei as classical particles, and quantum (PIMD) simulations of the concerted proton transfer between amine and imine groups within the chain at 20 K and 10 K, showing significant differences with respect to the classical free-energy profile at 10 K. The height of the quantum free-energy profile at 20 K decreases by approximately half with respect to the classical barrier and it is further lowered at 10 K. Moreover, the shape of the barrier also changes significantly, showing double well character with a concave dip in the central part of the barrier at 10 K, see Fig. 3a. This demonstrates that the NQEs cause strong proton delocalization across the tunneling barrier, revealing the presence of deep proton tunneling, as shown in Fig. 3b.

According to the quantum simulations, the intermolecular distance between two nitrogen atoms (3.03 Å) decreases with respect to the classical case (3.12 Å), facilitating the proton transfer. Moreover, NQEs not only change the spatial redistribution of hydrogens but also the adjacent nitrogen atoms, creating the symmetric atomic arrangement as shown on Fig. 3c. This symmetric atomic arrangement not only explains the observed AFM contrast of the chains (see Fig. 2e), but also facilitates more direct interaction between proton and nitrogen atoms, which alongside with the shortening of N-N bonds enhances the electrostatic interaction[5]. These effects may partially explain the experimentally observed mechanical stability of the symmetric chains. In addition, PIMD simulations with deuterated DABQDI molecules showed a substantial increase of the activation barrier with only limited proton tunneling (see Fig. S14).

To understand a detailed mechanism of the proton tunneling process, we analyzed the correlation between positions of individual protons during the tunneling transfer. Figure 3e shows the spatial distribution of δ-reaction coordinates of two selected hydrogen bonds against the average δ-coordinate of all the hydrogen bonds in the chain (for the definition of δ-reaction coordinates see Fig. 3d). The δ-coordinate distribution plots show two well-localized peaks, which correspond to the two isomeric π-conjugations. The particular diagonal position of the distribution peaks reveals the concerted motion of all the protons during the tunneling process. This concerted motion can be rationalized as a preservation of the appropriate π-conjugation of the systems, which would be violated by an asynchronous proton transfer that would consequently increase the total energy of the system. In contrast, pure proton transfer would again violate the π-conjugation, i.e. alternation of single and double bonds. Thus, proton transfer has to be accompanied by electronic rearrangement, and we can interpret the process as proton-coupled electron transfer[40] introducing a strong coupling between the electronic and vibrational proton degrees of freedom. This can explain why our adiabatic DFT and PIMD simulations are not able to reproduce the appearance of in-gap end states, as discussed next.

From this perspective, the ground state of the chain should be viewed as a linear combination of two isomeric π-conjugations, which may effectively lead to the delocalization of π electrons over the chain. We note that the symmetric chains are characterized by the presence of bright spots at their ends, visible in STM imaging (Fig. 1b). To understand the origin of the bright spots, we performed point scanning tunneling spectroscopy (STS; see Methods) combined with high-resolution AFM imaging. This allows us to unambiguously correlate the chemical and electronic structure of the chain ends. Point STS revealing the electronic structure of the end state are shown in Fig. 4a) for the positions indicated in Fig. 4b). Similar to the chain interior, the AFM image (Fig. 4b) shows a highly symmetric contrast supporting the presence of proton tunneling between two terminal nitrogen atoms. Figure 4d shows 1D STS spectroscopies taken along the central symmetry axis of the chain end (red dashed line in Fig. 4c, see also Fig. S9), revealing the presence of an in-gap state tightly localized around the terminating molecule unit. The end-state is centered at 50 meV above the Fermi level. For the sake of clarity, selected spectra from spatial coordinates indicated in Fig. 4b are shown in Fig. 4a. This state only appears on the symmetric chain termination, remaining completely absent in the case of canted chains (see Fig. S3 a, b). Thus, we attribute the appearance of the end-states (ESs) in the symmetric chains to the concerted proton tunneling motion, which allows the π-electron system to be effectively delocalized across the whole chain.

To understand the origin of the in-gap ESs, we analyzed the electronic structure of a tight-binding Hamiltonian mapping π-conjugated electrons of the chain (for detailed description see SI) as illustrated in Fig. 4e. First, we extracted tight-binding parameters representing π-electrons of an isolated molecule from our local basis DFT calculations[35] using the Löwdin transformation of exclusively $p_z$-orbitals (the electronic spectrum of single molecule is shown in Fig. 4f, left). We also introduced an intermolecular hopping $\tau_{hbond}$ between molecular units to the Hamiltonian (Fig. 4e), which in the case of classical H-bonds is much smaller ($\tau_{hbond} < 0.1eV$) than the hopping corresponding to covalent double or single bonds. Consequently, the electronic structure of a chain consisting of only weakly electronically coupled molecular units remains very similar to the single molecule case with only a slight broadening of the molecular levels into bands (see Fig. 4f). On the other hand, the presence of concerted proton tunneling with its strong proton-coupled electron transfer, enhances the electronic coupling of the π-electrons in the system. To consider these effects we introduced an extra hopping $\tau_{edge}$ between two nitrogen atoms at the edges of the chain, shown on Fig. 4e. We have analysed the influence of the parameters $\tau_{hbond}$ and $\tau_{edge}$ on the electronic spectrum of this model Hamiltonian. For certain values of the hopping parameters $\tau_{hbond}$ and $\tau_{edge}$ corresponding to weak covalent bonds (slightly weaker than intramolecular covalent bonds formed by π-electrons of carbon and nitrogen atoms), we find in-gap states (the electronic spectra shown in Fig. 4f, with in-gap states highlighted by colors) spatially localized at the edges of the chain (Fig. 4h right) in contrast to the bulk-like states that are delocalized over the whole chain length (Fig. 4g right). Note that the enhancement of the intermolecular hopping $\tau_{hbond}$ causes a reduction of the band gap of the straight chains relative to canted, which is in good agreement with STS measurements, see Fig. S10. In this way the experimental observation of the ESs and the renormalization of the band gap provides additional support to the enhanced intermolecular interaction picture, which underlies the large mechanical stability of the chains observed experimentally.

Our results make a direct connection between concerted proton tunneling and the two signal characteristics of the symmetric chains: their enhanced mechanical stability, and the appearance of in-gap end-states. Moreover, this work demonstrates that NQEs cause the emergence of electronic states at the Fermi level, a high proton tunneling rate and strong electron-proton coupling, which lead to the delocalization of π-electrons within the molecular chain. These phenomena are relevant to the ingredients used to establish the high-temperature superconducting state recently observed in high-pressure hydrates,[41,42] and it indicates that the strength of hydrogen bonds may be enhanced close to covalency[43] due to NQEs. We believe that these findings will stimulate further investigation of nuclear quantum phenomena including a search for similar systems beyond 1D where concerted proton motion and enhanced proton-coupled electron transfer strongly affect their collective mechanical and electronic properties.

# References

1. Benoit, M., Marx, D. & Parrinello, M. Tunnelling and zero-point motion in high-pressure ice. *Nature* **392**, 258–261 (1998).

2. Horiuchi, S. *et al.* Above-room-temperature ferroelectricity in a single-component molecular crystal. *Nature* **463**, 789–792 (2010).

3. Frank, R. A. W., Titman, C. M., Pratap, J. V., Luisi, B. F. & Perham, R. N. A molecular switch and proton wire synchronize the active sites in thiamine enzymes. *Science* **306**, 872–876 (2004).

4. Salna, B., Benabbas, A., Sage, J. T., Van Thor, J. & Champion, P. M. Wide-dynamic-range kinetic investigations of deep proton tunnelling in proteins. *Nat. Chem.* **8**, 874–880 (2016).

5. Li, X. Z., Walker, B. & Michaelides, A. Quantum nature of the hydrogen bond. *Proc. Natl. Acad. Sci. U. S. A.* **108**, 6369–6373 (2011).

6. Guo, J. *et al.* Nuclear quantum effects of hydrogen bonds probed by tip-enhanced inelastic electron tunneling. *Science* **352**, 321–325 (2016).

7. Wang, L., Fried, S. D., Boxer, S. G. & Markland, T. E. Quantum delocalization of protons in the hydrogen-bond network of an enzyme active site. *Proc. Natl. Acad. Sci. U. S. A.* **111**, 18454–18459 (2014).

8. Bove, L. E., Klotz, S., Paciaroni, A. & Sacchetti, F. Anomalous Proton Dynamics in Ice at Low Temperatures. *Phys. Rev. Lett.* **103**, 165901 (2009).

9. Lin, L., Morrone, J. A. & Car, R. Correlated Tunneling in Hydrogen Bonds. *J. Stat. Phys.* **145**, 365–384 (2011).

10. Drechsel-Grau, C. & Marx, D. Quantum simulation of collective proton tunneling in hexagonal ice crystals. *Phys. Rev. Lett.* **112**, 148302 (2014).

11. Gross, L. *et al.* The Chemical Structure of a Molecule Resolved by Atomic Force Microscopy. *Science* **325**, 1110–1114 (2009).

12. Jelinek, P. High resolution SPM imaging of organic molecules with functionalized tips. *J. Phys. Condens. Matter* **29**, 343002 (2017).

13. Meng, X. *et al.* Direct visualization of concerted proton tunnelling in a water nanocluster. *Nat. Phys.* **11**, 235–239 (2015).

14. Pascal, S. & Siri, O. Benzoquinonediimine ligands: Synthesis, coordination chemistry and

properties. *Coord. Chem. Rev.* **350**, 178–195 (2017).

15. Siri, O., Braunstein, P., Rohmer, M. M., Bénard, M. & Welter, R. Novel 'Potentially Antiaromatic', Acidichromic Quinonediimines with Tunable Delocalization of Their 6π-Electron Subunits. *J. Am. Chem. Soc.* **125**, 13793–13803 (2003).

16. Dähne, S. & Leupold, D. Coupling Principles in Organic Dyes. *Angew. Chemie Int. Ed. English* **5**, 984–993 (1966).

17. Rumpel, H. & Limbach, H. H. NMR Study of Kinetic HH/HD/DD Isotope, Solvent, and Solid-State Effects on the Double Proton Transfer in Azophenine. *J. Am. Chem. Soc.* **111**, 5429–5441 (1989).

18. Nietzki, R. & Hagenbach, E. Ueber Tetramidobenzol und seine Derivate. *Berichte der Dtsch. Chem. Gesellschaft* **20**, 328–338 (1887).

19. Audi, H. *et al.* Extendable nickel complex tapes that reach NIR absorptions. *Chem. Commun.* **50**, 15140–15143 (2014).

20. Griffiths, D. J. & Schroeter, D. F. *Introduction to Quantum Mechanics*. (Cambridge University Press, 2018). doi:10.1017/9781316995433

21. Gross, L. *et al.* Bond-order discrimination by atomic force microscopy. *Science* **337**, 1326–1329 (2012).

22. de Oteyza, D. G. *et al.* Direct Imaging of Covalent Bond Structure in Single-Molecule Chemical Reactions. *Science* **340**, 1434–1437 (2013).

23. Emmrich, M. *et al.* Subatomic resolution force microscopy reveals internal structure and adsorption sites of small iron clusters. *Science* **348**, 308–311 (2015).

24. Hapala, P. *et al.* Mapping the electrostatic force field of single molecules from high-resolution scanning probe images. *Nat. Commun.* **7**, 11560 (2016).

25. de la Torre, B. *et al.* Non-covalent control of spin-state in metal-organic complex by positioning on N-doped graphene. *Nat. Commun.* **9**, 2831 (2018).

26. Blum, V. *et al.* Ab initio molecular simulations with numeric atom-centered orbitals. *Comput. Phys. Commun.* **180**, 2175–2196 (2009).

27. Becke, A. D. A new mixing of Hartree-Fock and local density-functional theories. *J. Chem. Phys.* **98**, 1372–1377 (1993).

28. Stephens, P. J., Devlin, F. J., Chabalowski, C. F. & Frisch, M. J. Ab Initio Calculation of

Vibrational Absorption and Circular Dichroism Spectra Using Density Functional Force Fields. *J. Phys. Chem.* **98**, 11623–11627 (1994).

29. Tkatchenko, A. & Scheffler, M. Accurate molecular van der Waals interactions from ground-state electron density and free-atom reference data. *Phys. Rev. Lett.* **102**, 073005 (2009).

30. Hapala, P. *et al.* Mechanism of high-resolution STM/AFM imaging with functionalized tips. *Phys. Rev. B* **90**, 085421 (2014).

31. Peng, J. *et al.* Weakly perturbative imaging of interfacial water with submolecular resolution by atomic force microscopy. *Nat. Commun.* **9**, 122 (2018).

32. Kapil, V. *et al.* i-PI 2.0: A universal force engine for advanced molecular simulations. *Comput. Phys. Commun.* **236**, 214–223 (2019).

33. Plimpton, S. Fast Parallel Algorithms for Short-Range Molecular Dynamics. *J. Comput. Phys.* **117**, 1–19 (1995).

34. Mendieta-Moreno, J. I. *et al.* Fireball / amber: An efficient local-orbital DFT QM/MM method for biomolecular systems. *J. Chem. Theory Comput.* **10**, 2185–2193 (2014).

35. Lewis, J. P. *et al.* Advances and applications in the FIREBALL ab initio tight-binding molecular-dynamics formalism. *Phys. Status Solidi Basic Res.* **248**, 1989–2007 (2011).

36. Lee, C., Yang, W. & Parr, R. G. Development of the Colle-Salvetti correlation-energy formula into a functional of the electron density. *Phys. Rev. B* **37**, 785–789 (1988).

37. Grimme, S., Ehrlich, S. & Goerigk, L. Effect of the damping function in dispersion corrected density functional theory. *J. Comput. Chem.* **32**, 1456–1465 (2011).

38. Heinz, H., Lin, T. J., Kishore Mishra, R. & Emami, F. S. Thermodynamically consistent force fields for the assembly of inorganic, organic, and biological nanostructures: The INTERFACE force field. *Langmuir* **29**, 1754–1765 (2013).

39. Kumar, S., Rosenberg, J. M., Bouzida, D., Swendsen, R. H. & Kollman, P. A. THE weighted histogram analysis method for free-energy calculations on biomolecules. I. The method. *J. Comput. Chem.* **13**, 1011–1021 (1992).

40. Huynh, M. H. V. & Meyer, T. J. Proton-coupled electron transfer. *Chem. Rev.* **107**, 5004–5064 (2007).

41. Drozdov, A. P., Eremets, M. I., Troyan, I. A., Ksenofontov, V. & Shylin, S. I. Conventional superconductivity at 203 kelvin at high pressures in the sulfur hydride system. *Nature* **525**, 73–76

(2015).

42. Drozdov, A. P. *et al.* Superconductivity at 250 K in lanthanum hydride under high pressures. *Nature* **569**, 528–531 (2019).

43. Grabowski, S. J. What is the covalency of hydrogen bonding? *Chem. Rev.* **111**, 2597–2625 (2011).

## Acknowledgements

O.S. thanks the Centre National de la Recherche Scientifique (CNRS) and the Ministère de la Recherche et des Nouvelles Technologies. This work was supported by the Czech Science Foundation (Reg. No. 18-09914 S, 20-13692X), We acknowledge CzechNanoLab Research Infrastructure supported by MEYS CR (LM2018110). P.J. acknowledges support from Praemium Academie of the CAS. O.M. was supported by the Primus16/SCI/27/247019 grant from Charles University. Computational resources were provided by the CESNET LM2015042 and the CERIT Scientific Cloud LM2015085, provided under the programme "Projects of Large Research, Development, and Innovations Infrastructures".

## Contributions

P.J. and O.S. conceived the project and designed the experiments. A. C., J. H., V.M. S. and M. Š. performed and analyzed the SPM experiments. S. P. and O. S. synthesized organic molecules. J.I.M., D.S., S.I.E., K.V. and P.J. performed tight bonding modelling, P.M. performed DFT and AFM calculations, J.I.M., O.M. performed PIMD calculations. J.I.M., D.S., S.I.E., K.V, O.M. and P.J interpreted theoretical results and underlying mechanism. All authors discussed the results, co-wrote and commented on the manuscript.

These authors contributed equally: Aleš Cahlík, Jack Hellerstedt, Jesus I. Mendieta-Moreno.

## Competing interests

The authors declare no competing interests.

## Data availability

The data that support the findings of this study are available from the corresponding authors upon reasonable request.

# Figures

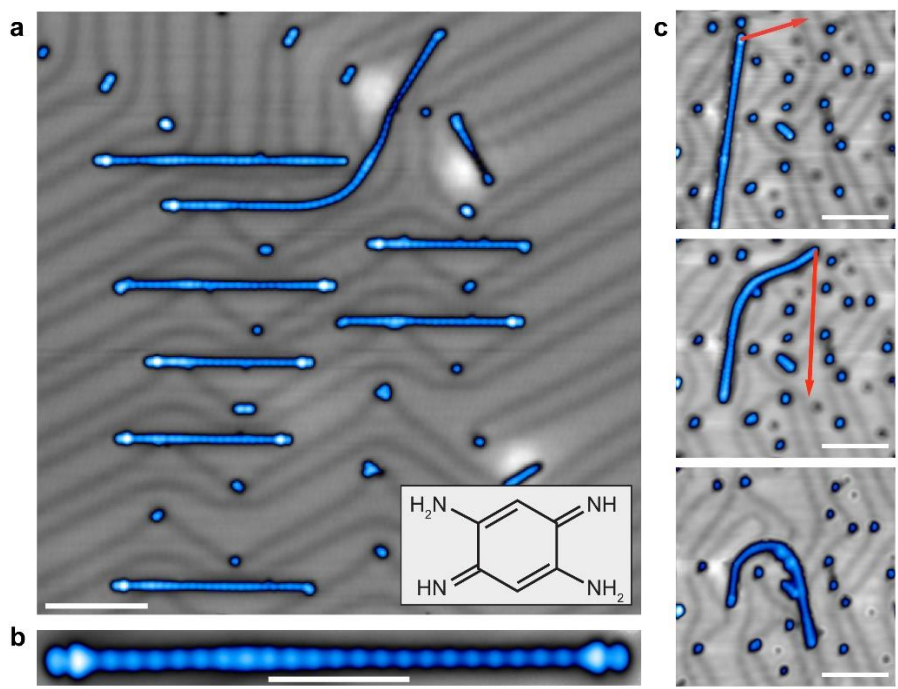

**Figure 1:** *a) Representative overview STM image of molecular chains and single molecule species. (50 mV, 10 pA, scale bar 10nm). Inset: 2,5-diamino-1,4-benzoquinonediimine (DABQDI) structure. b) Close-up STM image of the symmetric chain with characteristic bright spots at the ends. (30 mV, 5 pA, scale bar 5 nm) c) From top to bottom: sequentially acquired STM images of chain manipulation experiment. Red arrows represent the probe movement after contacting the chain end (procedure detailed in Methods) (all images 100 mV, 10 pA, scale bars 10 nm).*

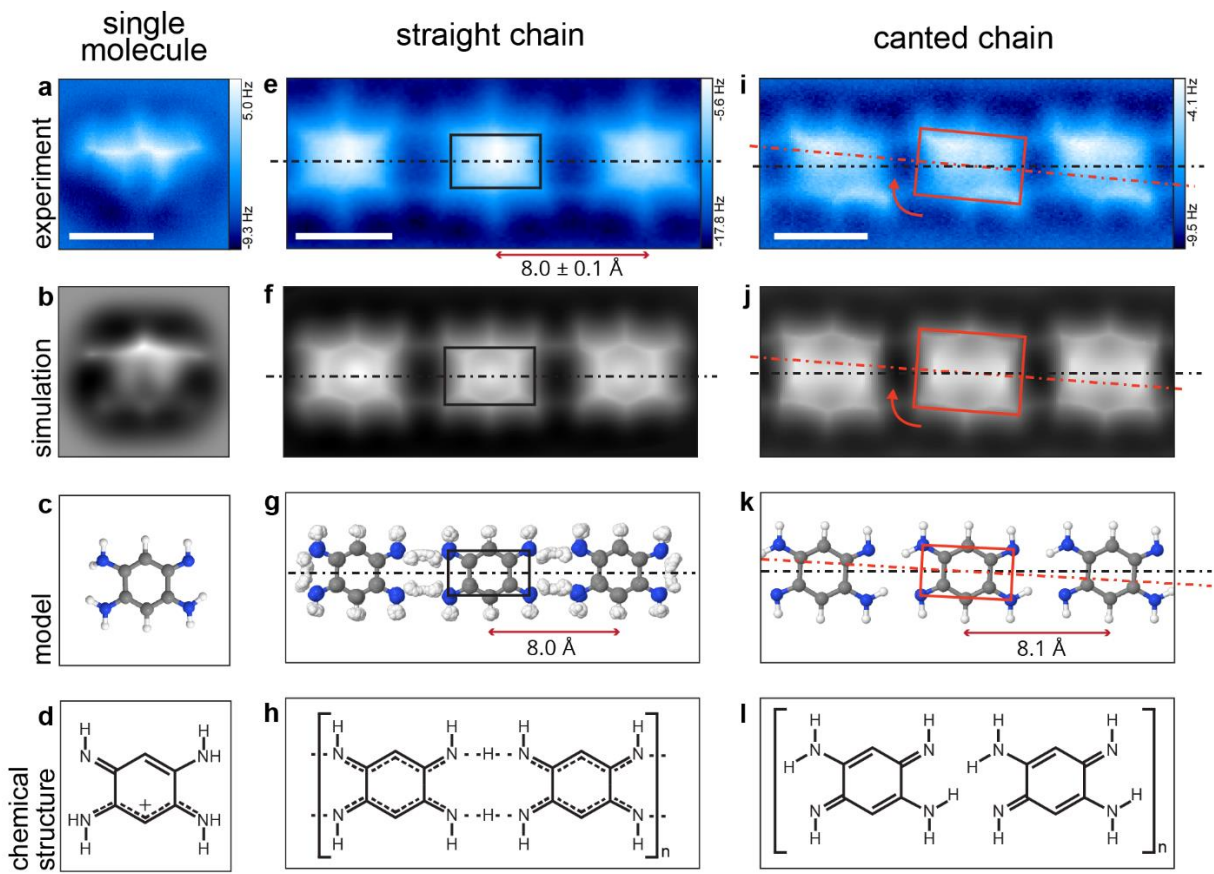

**Figure 2:** *Comparison between experimentally acquired high resolution AFM images of the observed molecular species and their respective simulated high resolution AFM images based on the calculated models (DFT, for c) and k)). a) - d) Single molecule with one additional proton, that impedes subsequent chain growth. e)- h) Hydrogen bonded symmetric chain with concerted proton tunneling. Molecular units are symmetric around the chain axis. Model (g) calculated by PIMD at the transition state. i)- l) Hydrogen bonded canted chain (no proton tunneling). Molecules are canted with respect to the main axis and distinct contrast difference between imine and amine groups is visible (All scale bars are 500 pm).*

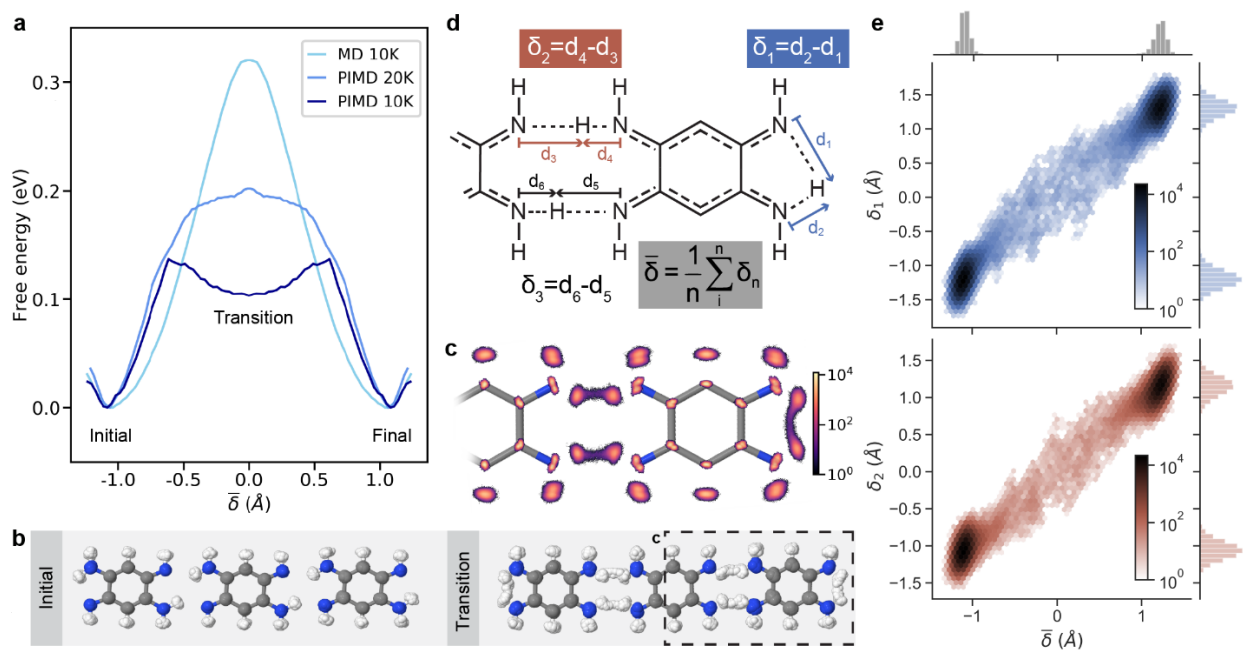

**Figure 3:** *a) Free energy curve of the proton transfer calculated using classical MD at 10 K (cyan), Path Integral MD at 10 K (light blue) and Path Integral MD at 20 K (dark blue) with the average $\bar{\delta}$ of all hydrogen bonds as the reaction coordinate. b) PIMD structure for $\delta = -1$ Å (initial) and $\delta = 0$ Å (transition) c) 2D histogram of atomic density projected in the plane of the molecules at the transition state. d) Scheme of δ reaction coordinates. e) Correlation between the average delta and $\delta_2$ (intra) and $\delta_1$ (edge) with 2D distribution in logarithmic scale and marginal distributions in linear scale, respectively.*

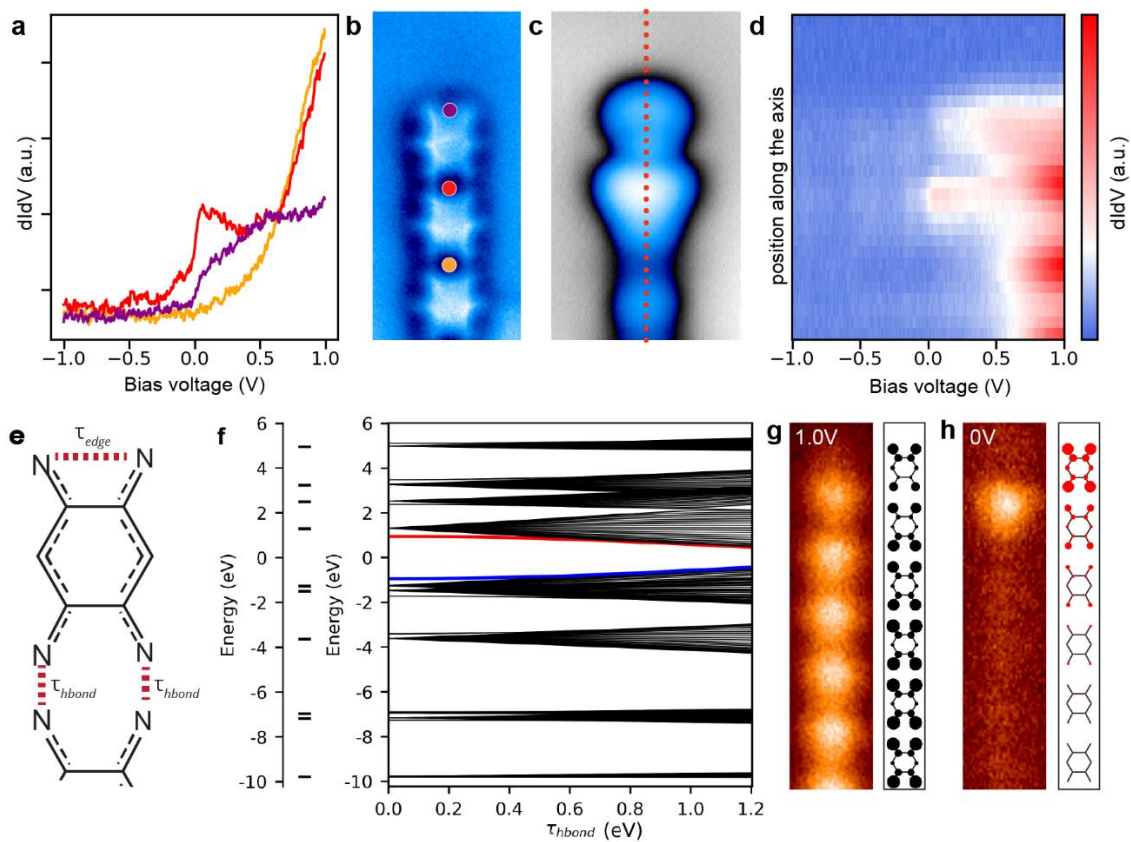

**Figure 4:** *a) Selected point STS spectra, taken at positions marked in b), showing the presence of the in gap electronic state. b) Representative experimental high resolution AFM image of the chain end. c) Representative experimental STM image of the chain end. (30mV, 10pA) d) Map of STS spectra taken along the chain axis at positions indicated in c) showing the spatial localization of the end state around the terminating molecular unit. e) Definition of coupling parameters $\tau_{hbond}$ and $\tau_{edge}$ in the tight-binding model of a molecular chain (only non-hydrogen atoms are shown as they are the sole hosts of $p_z$ orbitals). f). The electronic spectra of single-molecule (left) and molecular chain (right, model composed of 20 molecules) as function of parameter $\tau_{hbond}$ ($\tau_{edge}$ = 0.9 eV). In-gap states highlighted in red and blue. g), h) Spatially-resolved differential conductance maps (left) acquired above the last six molecules of a straight chain at 1V and 0V respectively demonstrate the localization of the in-gap state. Density of states of the tight-binding model projected to every atom (right) for a sum of ten delocalized bulk-like states (DS, black) and the in-gap state (ES, orange). For a better comparison with the experimental images, only the last six molecules are shown.*

## Materials and methods

### Chain growth

The precursor molecule 2,5-diamino-1,4-benzoquinonediimines (DABQDI) was synthesized via the procedure described in the literature[19]. Atomically clean Au(111) was prepared by repeated Argon ion (Ar$^+$) sputtering/annealing cycles. Molecules were sublimed from a home-built evaporator with tantalum pocket onto the clean Au(111). To form **the straight chains**, three steps were followed: i) deposition of the DABQDI molecular precursor (crucible temperature 90-100°C) into the microscope head on a cold sample (5 K), ii) transfer of the sample out of the microscope head to the microscope chamber, where the sample was kept for 60 minutes on the wobble stick until it warmed up almost to the room temperature, iii) transfer of the sample back to the microscope head (5 K). To form **the canted chains**, molecules were evaporated in the preparation chamber (crucible temperature 90-120°C) onto a sample thermalized to 20-60°C before being immediately transferred to the microscope head. For more details please see the discussion in the Supplementary Material.

### STM/AFM measurements

All experiments were performed in commercial ultrahigh vacuum (UHV) low-temperature microscopes with combined STM/AFM capabilities (Specs-JT Kolibri: PtIr tip, $f_0$ = 1 MHz, Q = 120 k, K = 540 kN/m and Createc-qPlus: PtIr tip, $f_0$ = 30 kHz, Q = 50k, K = 1.8 kN/m). To manipulate the chains, the metallic tip was approached to the chain end at $V_{bias}$ = 5 mV until a characteristic, abrupt change in the current channel was observed. To achieve sub-molecular resolution, the tip apex was functionalized with a CO molecule lifted from the Au(111) substrate[11]. All STS data were acquired in constant height mode (open feedback loop) using the lock-in technique (Nanonis internal) with a bias modulation amplitude of 5 mV and frequency 932 Hz. Prior to STS data acquisition, the tips were calibrated with reference to the Au(111) Shockley surface state.

### DFT Calculations

Density functional theory (DFT) calculations were performed using the FHI-AIMS code[26] within exchange-correlation functional B3LYP[27,28] to describe the electronic properties of the gas-phase DABQDI molecule and of its protonated form adsorbed on the Au(111) substrate using a 6x6 unit cell. In all the calculations, we employed the tight settings for the atomic basic sets. The atomic structures were thoroughly relaxed until the *Hellman*-Feynman forces were smaller than $10^{-3}$ eVÅ$^{-1}$. We have used the Tkatchenko-Scheffler correction[29] to include van der Waals interactions in the calculations. Only the Γ-point was used for integration in the Brillouin zone.

**PIMD calculations**

All the simulations were performed with 3 quinone molecules in local orbital DFT with local basis set Fireball code[35]; the surface was simulated using the interface forcefield[38]. DFT Fireball calculations used the BLYP exchange-correlation functional[27,36] with D3 corrections[37]. Classical MD was performed using the QM/MM method Fireball/Amber[34], while PIMD was performed using the i-PI software[32] with QM/MM interactions calculated by Fireball and LAMMPS[33]. We have used 512 PIMD replicas at 20 K and 1024 replicas at 10 K. To see the convergence with the number of replicas, see the supplementary material.

For the PIMD QM/MM simulation, an initial minimization of 10000 steps was performed followed by a classical QM/MM of 20 000 steps with a time step of 0.5 fs. For the PIMD we started with the results of the classical QM/MM and performed 20 000 steps with a time step of 0.25 fs.

To obtain the free energy profile we performed umbrella sampling with the bias applied to the reaction coordinate of the path integral centroid configuration at 20 K and of two contracted replicas at 10 K. The free energy profile was generated using the WHAM method[39] with 5000 steps in each window and a bias force of 200 kcal/mol on the reaction coordinate.

**AFM simulations**

The AFM images were calculated using the probe particle model[30]. The parameters of the tip were chosen to mimic a CO-tip, using a quadrupole charge moment of -0.1 $e \cdot Å^2$,[31] and the lateral stiffness of the CO molecule set to 0.25 Nm$^{-1}$. The electrostatic interaction was described in the AFM calculations using the potential calculated by DFT. To simulate the probe dynamics we used typical values of a qPlus sensor, oscillation amplitude $A = 100$ pm, sensor stiffness k = 3600 N/m and eigenfrequency $f_0 = 30$ kHz. The simulated AFM image, shown in Fig. 2f, is calculated as an average of AFM images of each replica of the PIMD simulations at 10 K.

# Supplementary Information

**Analysis of single molecules on a gold surface**

To understand the chemical structure of single molecules frequently observed on a Au(111) surface, which do not participate in the formation of the chains, we have carried out a comparative study of experimental and simulated high-resolution AFM images of possible candidates on a Au(111) surface. As shown by Siri et al.[15] hydrogenation of benzoquinonediimine ligands (see Fig. S1a) results in delocalization of one of the formerly localized 6π-electron conjugated systems. Therefore, we considered three possible models: i) a hydrogenated DABQDI (Fig. S1b); ii) an intact DABQDI (Fig. S1c) and iii) a dehydrogenated DABQDI molecule (Fig. S1d) The simulated AFM images were calculated with the PP-AFM toolkit using fully optimized structures deposited on a Au(111) surface obtained from total energy DFT simulations.

Fig. S1b shows the optimized structure of a hydrogenated precursor and the simulated AFM images at two different tip-sample heights. The simulated AFM images agree very well with the experimental images of single molecular species observed on the sample surface (Fig. 2a). The asymmetry in the nc-AFM signal contrast arises due to the adsorption height difference (side view on right panel of Fig. S1b). We consider two possible sources of the additional hydrogen – either a hydrogen transferred from other molecules or residual hydrogen gas in the UHV system. In the STM overview images, we have observed step edges decorated with clusters of undistinguishable molecular residues. These could potentially be dehydrogenated molecular units; however, we were not able to fully experimentally confirm either of these scenarios.

On the other hand, the calculated AFM images of a single DABQDI molecule fit well to the experimental AFM images acquired after deposition of DABQDI precursors onto the Au(111) substrate held at 5 K. (see Fig. S1c). In the case of the dehydrogenated DABQDI molecule Fig. S1d), we could not find any experimental counterparts.

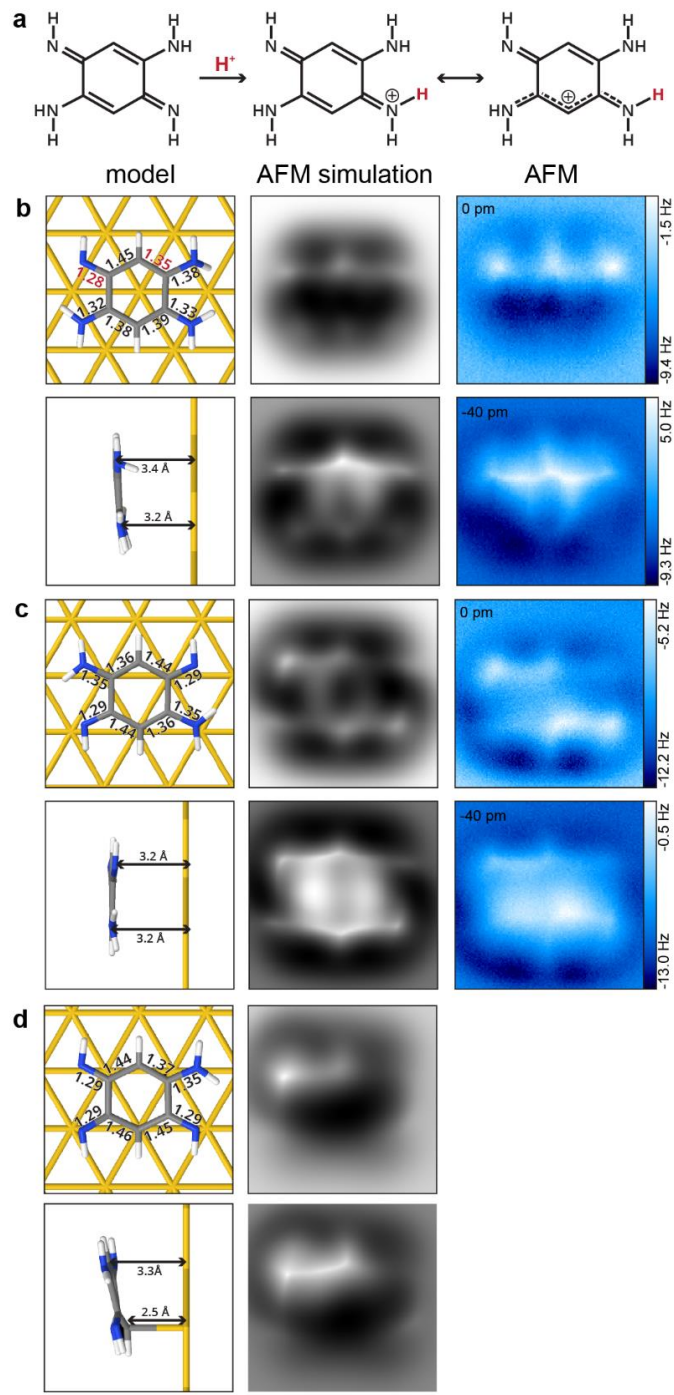

**Figure S1:** *a) Protonation reaction of 2,5-diamino-1,4-benzoquinonediimine. b) DFT calculated structure, simulated AFM images and assigned experimental AFM images acquired at two distinct tip heights of $C_6N_4H_9$ (deposition on a sample kept at RT). Comparison of interatomic distances in the model confirms the delocalization of the protonated conjugated system. The contrast asymmetry in the AFM images stems from the adsorption height difference. c) DFT calculated structure, simulated AFM images and assigned experimental AFM images acquired at two distinct tip heights of $C_6N_4H_8$ (deposition on a cold sample - 4 K). In this case the AFM image exhibits symmetric contrast. d) DFT calculated structure and simulated AFM image of $C_6N_4H_7$. The species was never experimentally observed.*

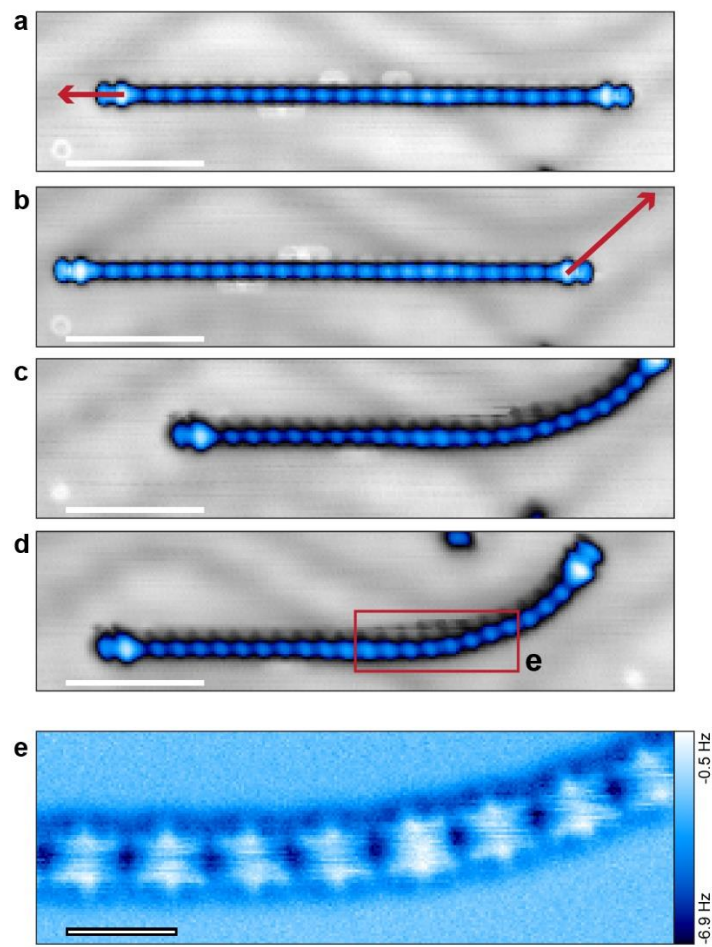

**Figure S2:** *a)-d) From top to bottom: sequentially acquired STM images of a manipulation experiment with straight chain. Red arrows represent the probe movement after contacting the chain end (procedure detailed in Methods). End states remain intact during all manipulation steps. (all images 5 mV, 10 pA, scale bars 5 nm) e) Experimentally acquired high resolution AFM image of the curved part (red rectangle in image d)) of straight chain after the last manipulation (scale bar 1nm).*

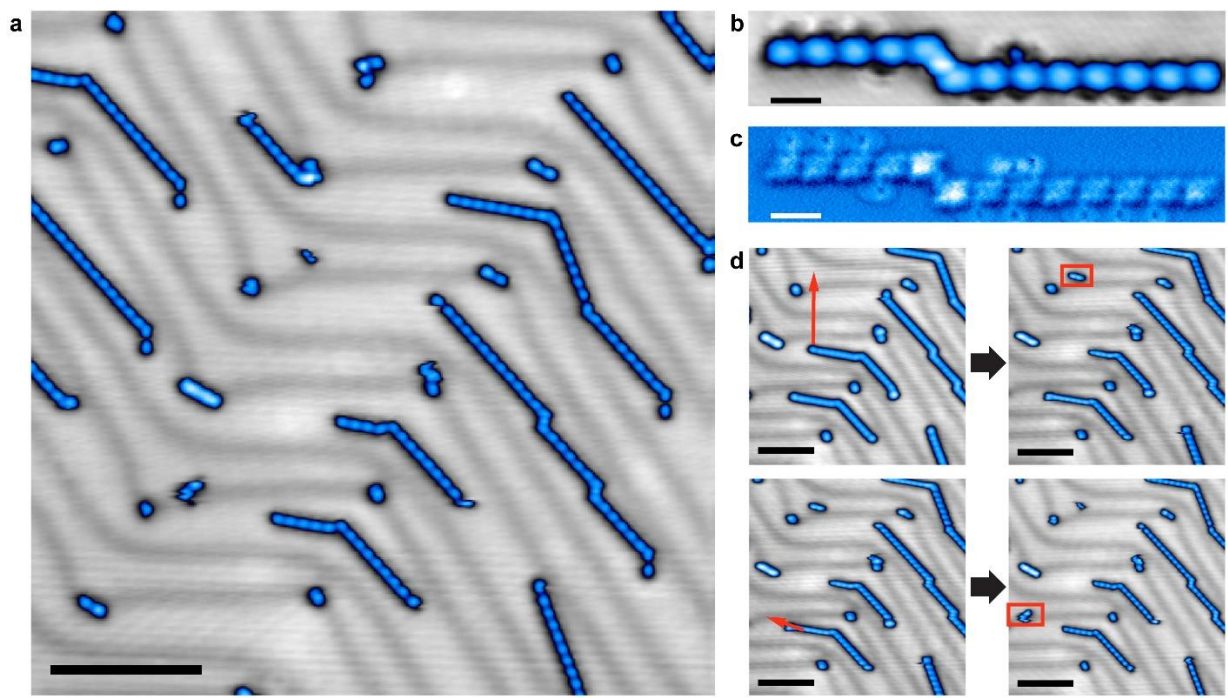

**Figure S3:** *a) Representative overview STM image of canted molecular chains. (140 mV, 20 pA, scale bar 10 nm) b) Close-up STM image of canted chain showing the absence of bright spots at the chain ends (140 mV, 20 pA, scale bar 1 nm). c) High resolution AFM image of the same chain showing its inner structure with visibly canted molecular units. d) STM images (140 mV, 20 pA) taken before and after two manipulation experiments. Red arrows represent the probe movement after contacting the chain end (procedure detailed in Methods). Images on the right side (after the manipulation) show apparent splitting of the chain into segments. Scale bars 10 nm.*

**Supporting arguments to exclude metal-organic Au-DABQDI as the symmetric chains**

Here we provide a series of arguments to rule out the presence of gold atoms in the linear chains.

1. Observation of defective chains with incorporated gold atoms

Fig. S5 shows an AFM image of a short defective chain, which we very rarely observe upon preparation of the sample at RT. The submolecular AFM contrast of such chains is quite distinct from those observed for the symmetric and canted chains. In the intermolecular region connecting two molecular units, a characteristic "x-like" feature is observed (marked by red arrows in the high pass filtered image shown in Fig. S5b). This feature fits well to our simulated AFM images of a molecular trimer with four-fold coordinated gold atoms between the molecular units on a Au(111) surface (see Fig. S5c). The total energy DFT calculations reveal only a minor out-of-molecular plane relaxation of four-fold coordinated gold atoms towards the surface, as shown in Fig. S5d. As was recently shown[44], a similar characteristic contrast can be observed in AFM images of gold porphyrins formed by on surface self-metalation of 2H-4FPP on a Au(111) surface.

Note that the simulated AFM image (Fig. S5c) shows slightly asymmetric contrast at both ends of the trimer. We tentatively attribute this experimentally observed symmetric contrast to the proton tunneling between the two nitrogen atoms at the end, which cannot be captured by classical DFT calculations.

2. Incommensurability of the linear chain with the Au(111) substrate

It could be argued that the DFT calculations fail to predict the correct optimal structure of the Au-DABQDI chains on the Au(111) surface. A strong out-of-plane displacement of gold atoms, resulting in their invisibility in AFM images, would indicate a strong interaction of gold atoms with the underlying substrate. Such a strong adatom-substrate interaction would require commensurability of the chain with the substrate. Fig. S4 shows a registration of the symmetric DABQDI chain with the Au(111) substrate using high-resolution AFM imaging, which clearly demonstrates the incommensurable alignment of the symmetric chain with the Au(111) substrate. Note that a strong interaction of the gold atoms with the substrate would also inhibit the possibility to readily manipulate the chains laterally on the surface with the probe.

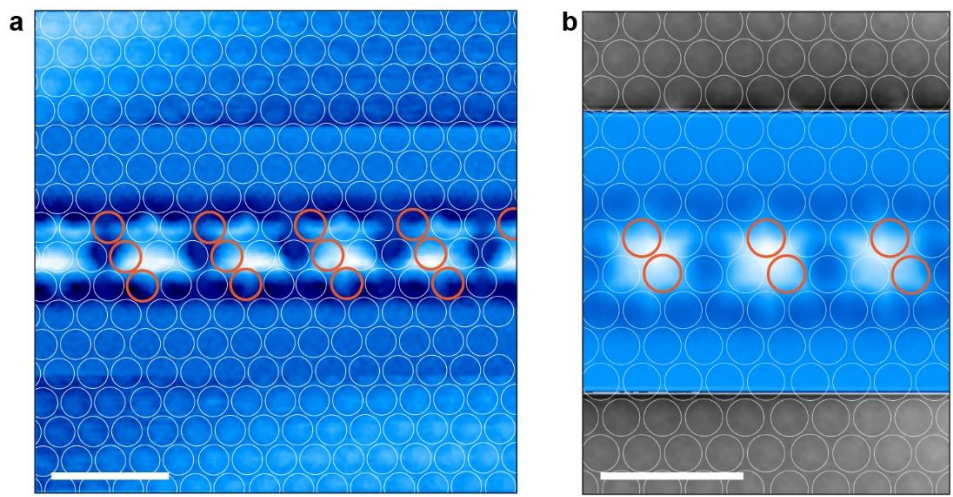

**Figure S4:** *a), b) Two experimentally acquired high resolution AFM images displaying the registration of the straight chain with the substrate demonstrating incommensurability with the underlying Au(111). Orange circles mark every third Au row to guide the eye. (scale bars 10nm)*

### 3. Formation of transition metal metal-organic chains at elevated temperatures

Thermodynamic considerations provide indirect evidence against the inclusion of Au adatoms in the chains of interest. We are able to grow 1D metal-organic chains up to hundreds of nanometers in length on Au(111) substrate by co-deposition of DABQDI and selected transition metals (TM = Fe, Co, Ni, Cr). In such chains, TM atoms adopt a four-fold coordination with the ligand similar to that shown for gold on Fig. S5d (separate manuscript under submission). A crucial parameter for the formation of the well-ordered and hundreds of nanometers long metal-organic chains on Au(111) surface is the co-deposition at elevated sample temperatures ~ 300ºC. For comparison we attach a STM overview and AFM image of these Fe-quinone chains (see Fig. S7a, b, unpublished data). However, if we only deposit the DABQDI itself on Au(111) surface in such elevated temperatures, we don't observe formation of any long-ordered metal-organic chains (the substrate remains clean). This indicates that the formation of long metal-organic Au-DABQDI is not energetically favorable even in such elevated temperatures. Consequently, the formation of well-ordered metal-organic Au-DABQDI chains at RT seems to be very unlikely. This also explains why we find very rarely those defective chains shown in Fig. S5a.

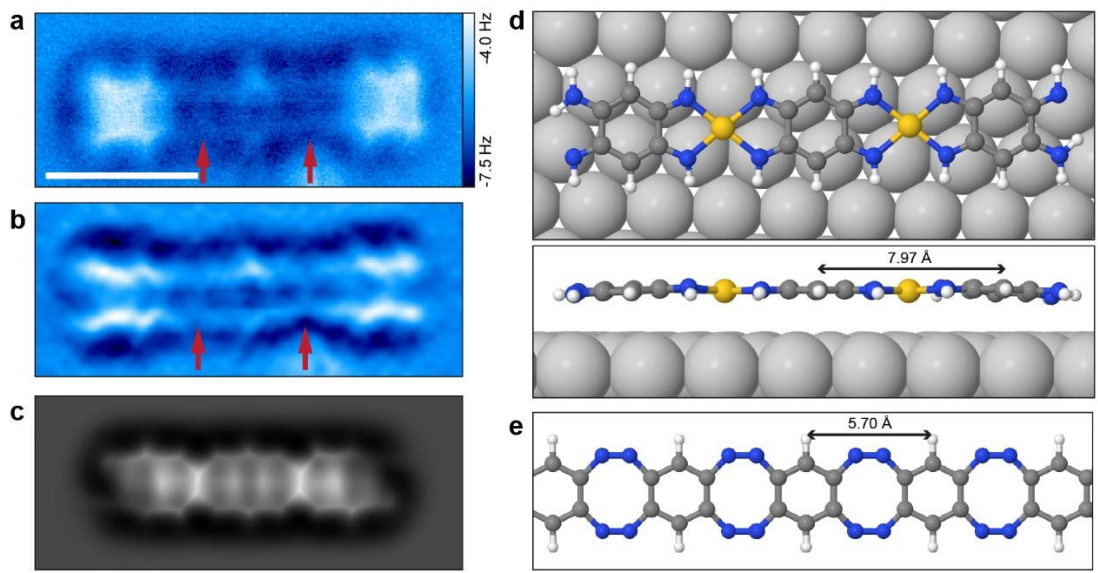

**Figure S5:** *a) Experimentally acquired high resolution AFM image of a trimer with incorporated Au adatoms. (scale bar 1 nm) b) High pass filtered version of AFM image a). Red arrows mark the characteristic "x-like" feature in between ligand units. c) Simulated high resolution AFM image based on the model shown in d). d) DFT calculated model of a trimer with incorporated Au adatoms on Au(111) surface - top and side view showing a minor out-of-molecular plane relaxation of gold atoms towards the surface. e) DFT calculated model of a chain composed of covalently bonded molecules with periodicity 5.7 Å, ~30% shorter than the Au coordinated structure.*

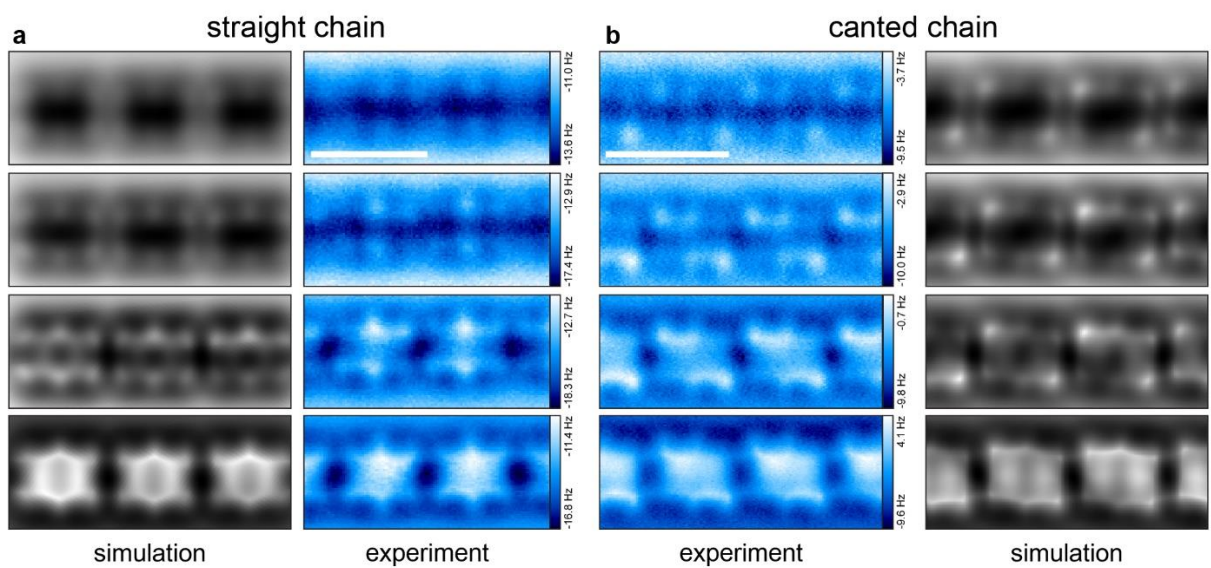

**Figure S6:** *Comparison between experimental AFM images of straight and canted chains acquired at different tip-heights and their respective simulated AFM images. (scale bars 1 nm)*

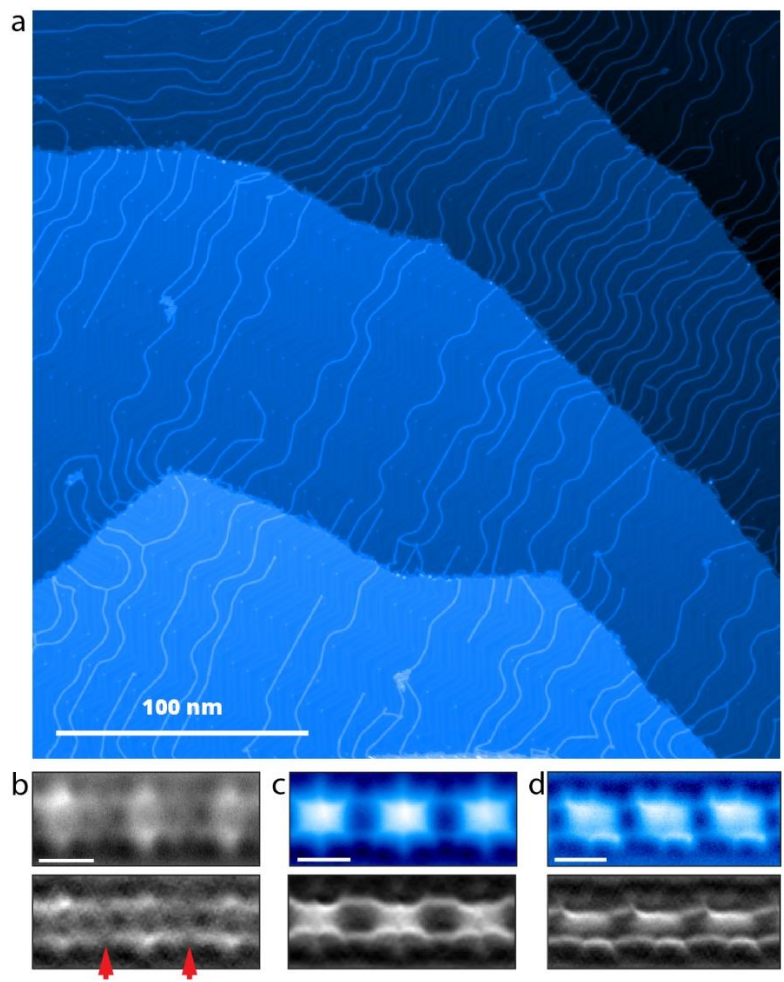

**Figure S7:** *a) Representative overview STM image of Fe-Quinone chains grown by a co-deposition of Fe and DABQDI molecules onto Au(111) substrate held at 400 ºC. b) Experimentally acquired high resolution AFM image and its high pass filtered version of the chain with incorporated Fe adatoms that exhibits similar "x-like" feature as shown in Fig. S5. c), d) Experimental AFM images of straight and canted chain respective for a direct comparison. (b-d) scale bars 500 pm)*

**Supporting arguments to exclude other possible scenarios**

With respect to other possible scenarios explaining the experimental observations, the UHV and low temperature environment make the possibility of any contaminant species very unlikely except for atomic hydrogen and gold atoms. Such well-defined conditions reduce the pool of possibilities significantly.

Besides the incorporation of gold adatoms, we explored the possibility of chains terminated with molecules with hydrogens either added or subtracted at the end. The presence of an extra hydrogen forming two amine $NH_2$ groups at the ends introduces strong steric repulsion between them, resulting in a non-planar structure that we do not observe with our ncAFM measurements. Both the non-planar and asymmetric arrangement is confirmed by our total energy DFT simulations, see Fig. S8a. We do not see any reason or mechanism explaining how the presence of the double amine groups at the ends of the chain would impose the fully symmetric arrangement of internal molecular units along the entire chain. Similarly, we can rule out a scenario including dehydrogenated ends of molecular chains, the total energy DFT simulations and related simulated AFM images shown in Fig. S8b. Furthermore, neither scenario could convincingly explain the enhanced mechanical stability of straight chains.

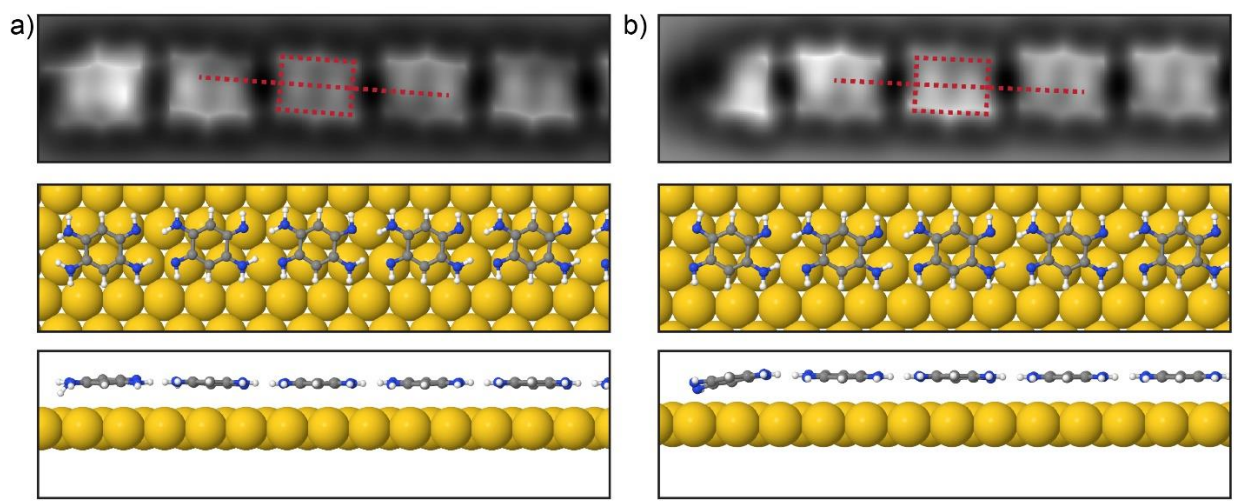

**Figure S8:** *Middle, Bottom: Top and side view of relaxed structure of a chain with the last molecule a) hydrogenated and b) dehydrogenated. Top: respective simulated AFM images. Molecules in the middle of the chain remain canted.*

**Notes on the sample preparation**

The procedure that reproducibly results in the formation of straight chains requires three steps:

- deposition of the DABQDI molecular precursor in **the microscope chamber** into the microscope head on a cold substrate (5 K) - single precursor molecules observed on the substrate (Fig. S1c),
- transfer of the sample out of the microscope head to the microscope chamber, where the sample is kept on the wobblestick until it warms up almost to the room temperature,
- transfer of the sample back to the microscope head (5 K) - straight chains are observed on the substrate.

In the case of a direct deposition of the molecular precursors in **the preparation chamber** onto the sample kept at RT, we have observed both the formation of straight and canted chains. Once preparation parameters for either type of growth (canted/straight) were established, we could reproducibly prepare several samples with the same type of chain growth. For the straight chain growth these parameters typically included lower sublimation temperature for precursor molecule. However, once we changed the experimental setup (a replacement of evaporator, new batch of molecules or change of the microscope system), the previously optimal parameters sometimes did not provide the desired chain growth anymore.

Despite our best effort we have not been able to reliably identify the crucial parameter unconditionally resulting in either chain growth. Therefore, to ensure the formation of the straight chains more reliably, we have tested the aforementioned procedure involving deposition on the substrate held at low temperature in the microscope head.

These observations suggest that the concerted proton tunneling is initiated by the conditions at which the molecules are deposited onto the surface at room temperature. One possible scenario is that the deposition on a sample at room temperature initiates the hydrogenation process of the precursor forming DABQDI species with an extra hydrogen, which we frequently find on the surface. The presence of critical amounts of defective molecules prevents formation of the straight chains.

**STS of the end states with different metallic tips**

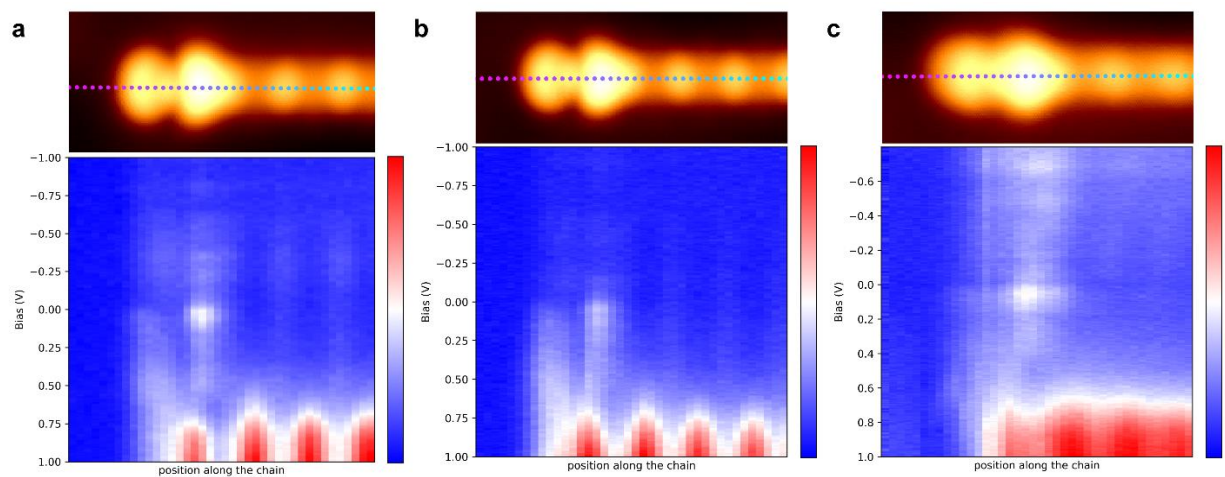

**Figure S9:** *a), b), c) Maps of STS spectra (bottom; dI/dV colorscale in a.u.) taken with different metallic tips along the chain axis at positions indicated in STM image (top) showing the spatial localization of the end state around the terminating molecular unit.*

**STS comparison of canted and straight chains**

Depending on the preparation conditions, in rare instances we have observed a formation of straight chains on a sample with predominantly canted chains. The dIdV spectra acquired over molecules in both types of chains with an identical tip are shown in Figure S10. From the position of LUMO orbital, it is apparent that the bandgap of canted chains is distinguishably larger than in the case of straight chains. Please note that we also observe instabilities in both kinds of chains when we try to reach critical biases of frontier orbitals, which induces chemical transformation of molecular units in the chain. For this reason, we were unable to reach the HOMO orbital in both types of chains, or reach states at higher positive bias.

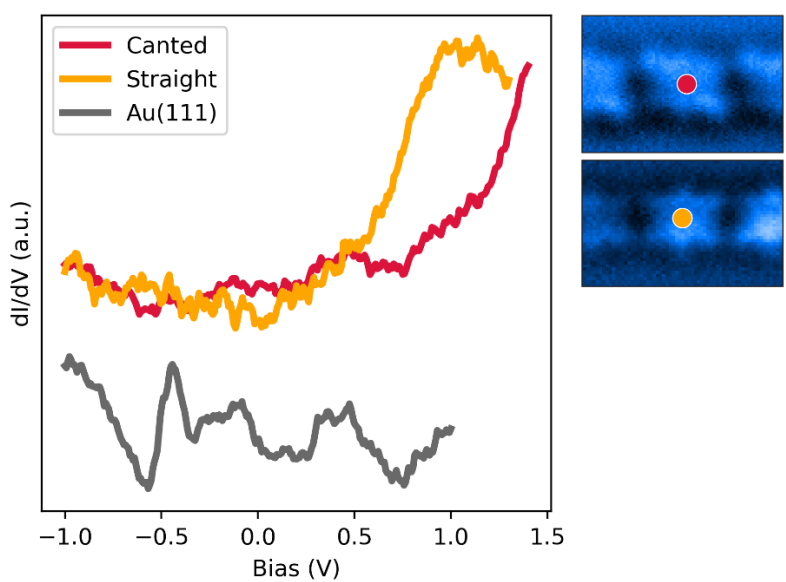

**Figure S10:** *Comparison of STS measurement acquired over DABQDI molecules in canted and straight chains recorded with the same tip. dIdV spectrum acquired over bare Au(111) surface is shown for reference.*

**The experiment with deuterated DABQDI precursors.**

There are two main obstacles to isolating the deuterated molecule in the solid state: 1/ the lack of solubility of unsubstituted DABQDI in chlorinated solvents (water free); and 2/ a back proton exchange of deuterated DABQDI molecules in the presence of hydrogen sources.

The H/D exchange is an equilibrium reaction in which the amount of deuterium should be high compared to the exchangeable protons of DABQDI. During the isolation process from a solvent (a mixture DMSO-d6 and CD3OD), the equilibrium is shifted towards the hydrogenated compound because of i) the presence of traces of water in these solvents, and ii) a decrease of the concentration of the deuterium source. This results in subsequent back proton exchange of deuterated DABQDI.

In this context, the need to use chlorinated solvents such as chloroform (CHCl3) is essential because of the extremely low concentration of water in these solvents (chlorinated solvents are immiscible with water in contrast to DMSO and methanol), preventing undesired side reactions (D/H exchange). Unfortunately, the lack of solubility of the unsubstituted DABQDI in such a solvent (as reported by Limbach and Rumpel[17]) impedes the deuteration reaction through this way.

Even if it was possible to prepare and isolate pure deuterated DABQDI, it is indispensable to use inert conditions in order to exclude air, moisture and other impurities that might catalyze back proton exchange (i.e. the exchange of deuterium with protons when there is a lack of a sufficiently large reservoir of deuterium). This presents a serious obstacle in the practicalities of introducing these molecules into the UHV microscope. During this procedure, we would unavoidably expose the deuterated molecules to the air and thus allow additional D/H exchange to occur. Thus, we do not see any option to ensure the controlled deposition of deuterated DABQDI precursor into the UHV chamber.

**Comparison of FHI-AIMS and Fireball DFT methods**

Path Integral Molecular Dynamics (PIMD) has been performed with the local orbital Fireball DFT package. To validate the accuracy of Fireball BLYP-D3 we compared it with the potential energy landscape of a single proton transfer in a periodic system calculated with the FHI-AIMS package using B3LYP-TS. Figure S11 compares the calculated energy barriers, which are similar. This justifies our use of the Fireball package to increase the conformational sample in our PIMD simulations.

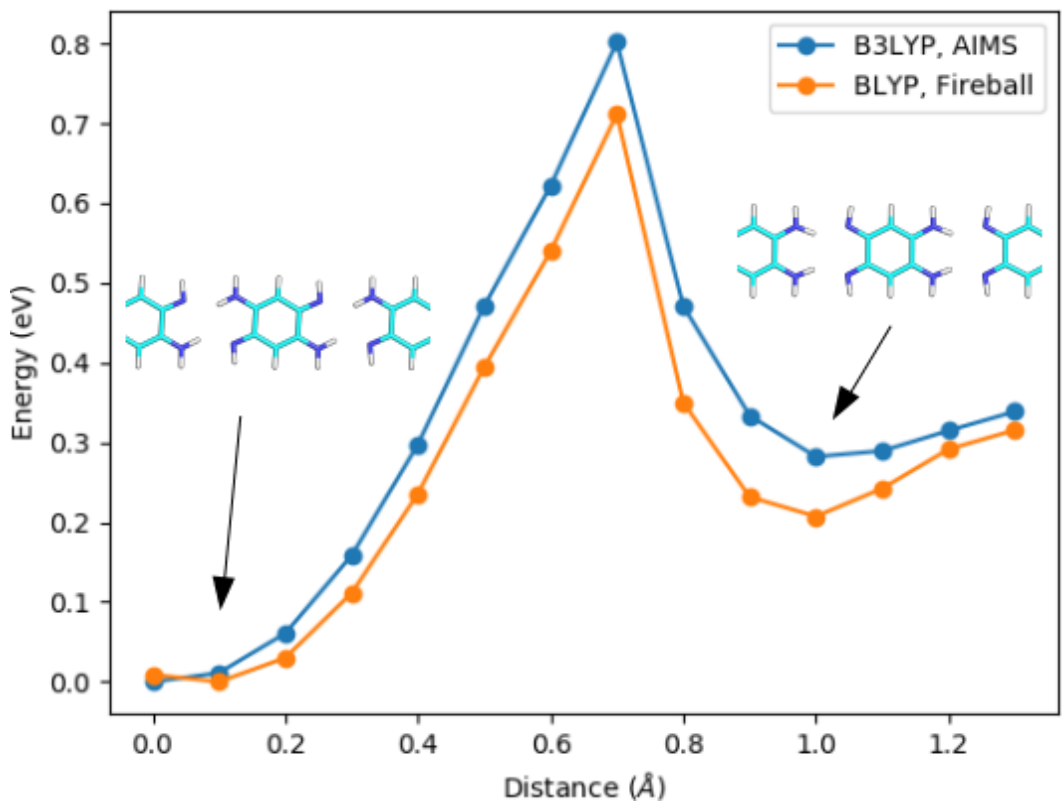

**Figure S11**: *Comparison of potential energy profiles of the single proton transfer in a periodic system obtained with FHI-AIMS B3LYP-TS (blue), and Fireball BLYP-D3 (orange) calculations.*

**Convergence based on the number of replicas in PIMD calculations at different temperatures**

The dependence of the number of replicas with temperature has been studied for 5, 10 and 20 K, with 32, 64, 128, 256, 512, 768 and 1024 replicas for the PIMD. To ensure that the system is stable we performed 10000 steps of PIMD with a time step of 0.25 fs for each pair of values. In order to see the convergence with the number of replicas for this system we represent the average over the last 2000 steps for each case, shown in Figure S12. Based on these simulations we conclude that we can do simulations at 20 K and 10 K with 1024 replicas. The simulations at 5 K seem to need more than 1024 replicas for full convergence and the computational cost to perform free energy calculations becomes prohibitive.

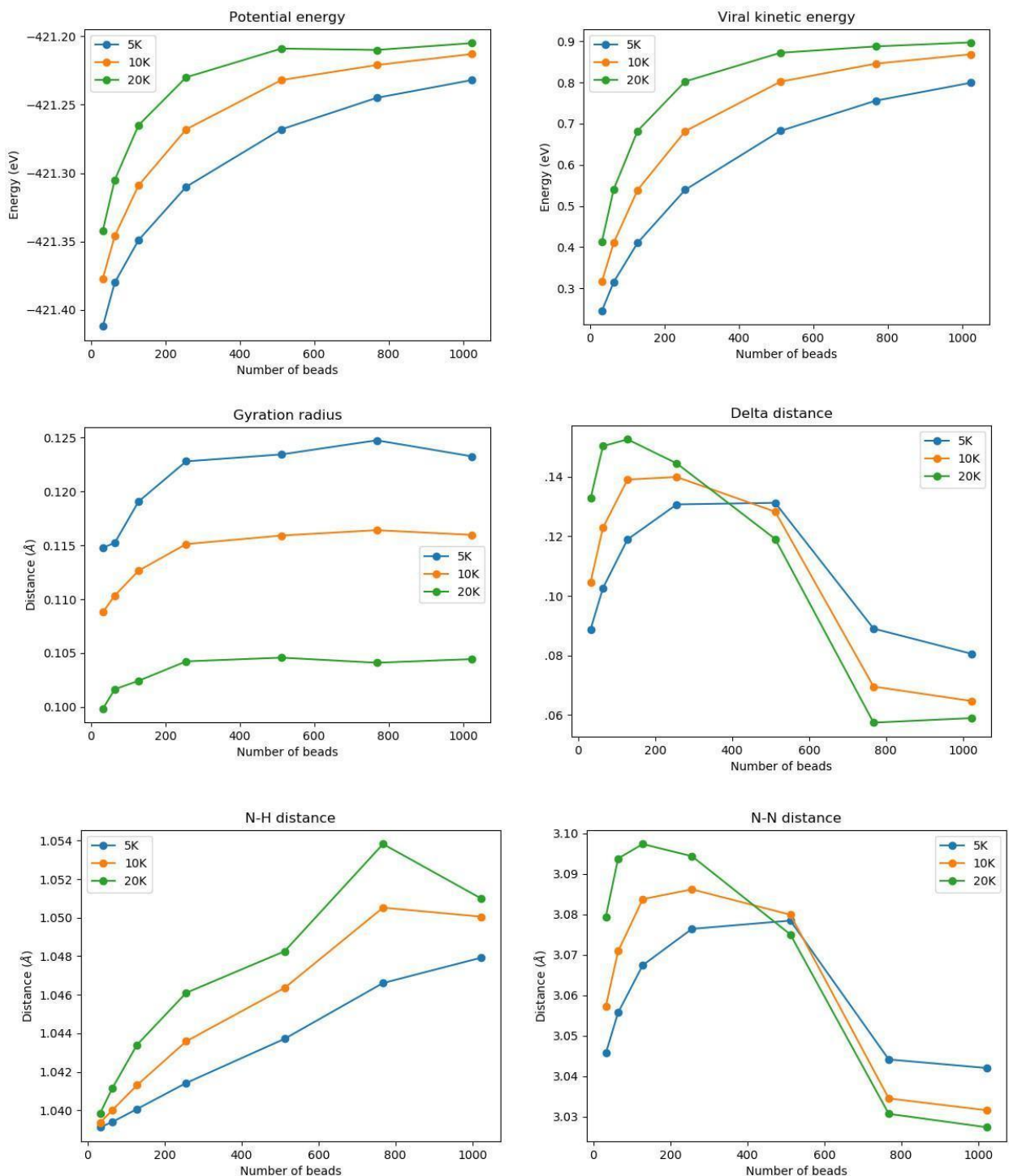

**Figure S12**: *Evolution of the mean value for the last 2000 steps of the PIMD for potential energy, viral kinetic energy, gyration radius delta distance, N-H distance and N-N distance at different temperatures (blue 5 K, green 10 K and red 20 K) when we change the number of replicas.*

**Free energy calculations**

Free energy profile calculations were carried out for a system consisting of three quinone molecules on the gold surface using the QM/MM scheme described below. We employed the WHAM method using 51 windows along the reaction coordinate. We have performed 3000 steps in each window with a time step of 0.25 fs.

To understand the behavior of the system, we considered distinct reaction coordinates transferring the hydrogen atoms from one nitrogen to the opposite nitrogen along the H-bond. Namely, we have tested as reaction coordinates the position of just one hydrogen atom, the collective reaction coordinate (average of the positions of all hydrogens), and also a different umbrella-like restraint for each hydrogen atom (see Figure S13).

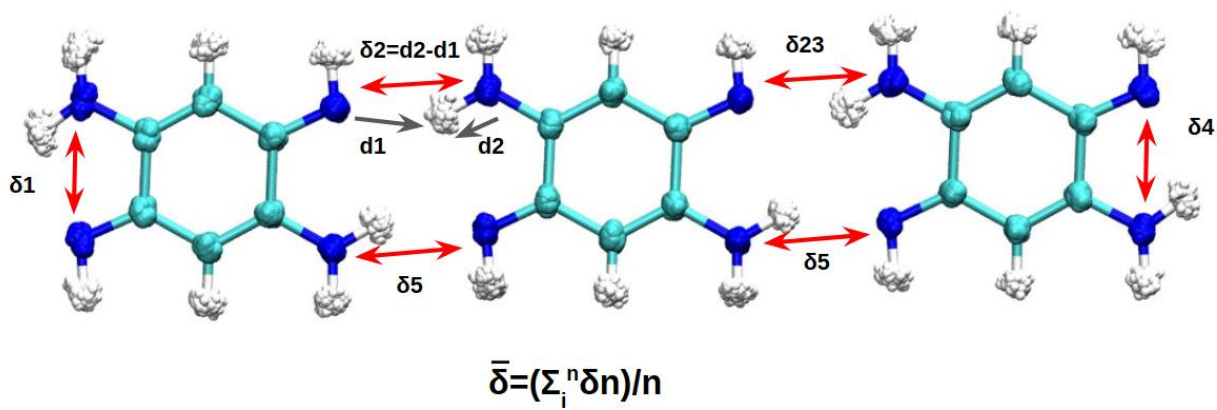

**Figure S13**: *Scheme of the reaction coordinates used for the classical and PIMD QM/MM simulations.*

The behavior of the system for the average of the $\delta$ and for six restraints is the same for PIMD simulations, consistent with the idea that coordinated proton tunneling takes place in the system. For the free energy profile at 20 K, a restraint has been added to the centroid of the ring polymer with 512 replicas. In the case of the 10 K free energy profile, a restraint has been added to two contracted replicas. This is due to the large delocalization of the ring polymer around $\delta=0$ Å, which creates hysteresis along the reaction when we apply the restraint only over the centroid.

For the classical QM/MM simulations, we have to carry out the simulation restraining every hydrogen simultaneously to obtain the same geometries that we observe in PIMD simulations. If we apply the restraint in the average $\delta$ in the classical MD simulations, we don't have a coordinated movement, but rather multiple barriers in a stepwise process. In Figure 3a of the main text the barrier for the classical MD at 10 K has been performed with a restraint on all the δ to be able to compare the energy.

We performed PIMD simulations replacing hydrogen with deuterium and found distinctive behavior of the deuterated chains. Figure S14 shows a comparison of the calculated activation barriers of concerted proton transfer for hydrogen and deuterium based DABQDI chains revealing that the activation energy is substantially increased - more than double and, in fact, comparable to the classical one, and that deuterium delocalization in the transient state is much smaller than in the case of hydrogen.

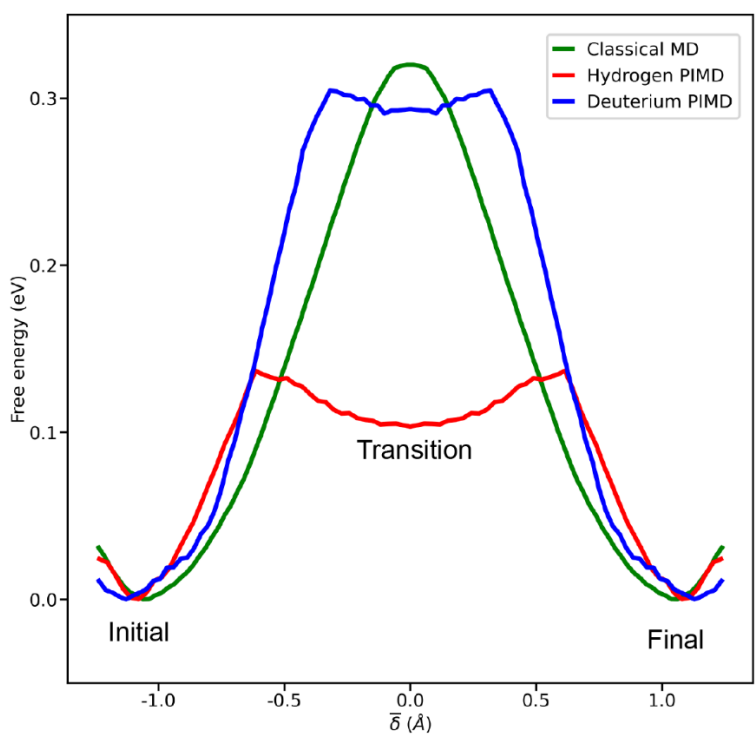

**Figure S14:** *Free energy profile at 10 K of the concerted proton transfer in DABQDI chains using classical MD(green) or path Integral MD for hydrogen(red) or deuterium(blue) for the transferring atoms. The delta reaction coordinate is defined in the main text. Note that the steeper increase of the free energy in the quantum case is due to the delocalization of the hydrogen/deuterium atom in the potential well.*

In Figure S15, appreciable differences between the MD and PIMD approaches can be observed. In the classical MD simulation, the protons are located just in the middle of the hydrogen bond, while in the PIMD simulations we observe a delocalization of the hydrogens along the hydrogen bond. This difference in behavior is responsible for the flatness of the barrier and characteristic of the deep tunneling regime. Moreover, we observe a further difference between the PIMD simulations at 10 K and 20 K.

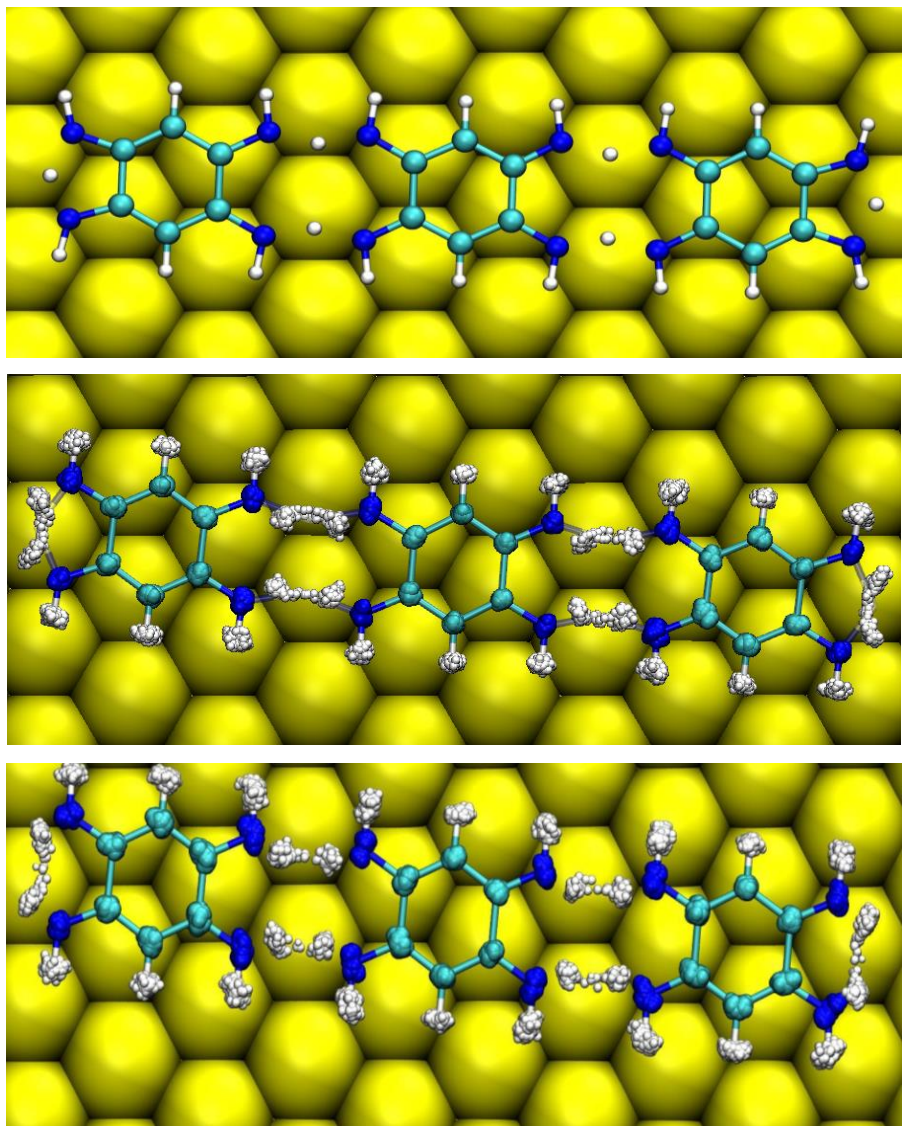

**Figure S15**: *Representation of the proton distribution at δ=0 Å with classical MD at 10 K (top), PIMD at 20 K (middle) and PIMD at 10 K (bottom).*

**Analysis of the concerted motion of protons**

To study the correlation of the different protons we perform a 5000 step of free dynamics calculation at 10 K starting in the window corresponding to δ=0 Å in the free energy profile. During the last 2000 steps of this PIMD we measured the δ coordinate of all the hydrogen bonds for the centroid trajectory and for all the 1024 replicas of our system. During this simulation the average δ stays around 0 Å but we can see that the distribution of the delta value for the replicas' trajectories has two clear peaks for each protonation state, as shown in Figure S16.

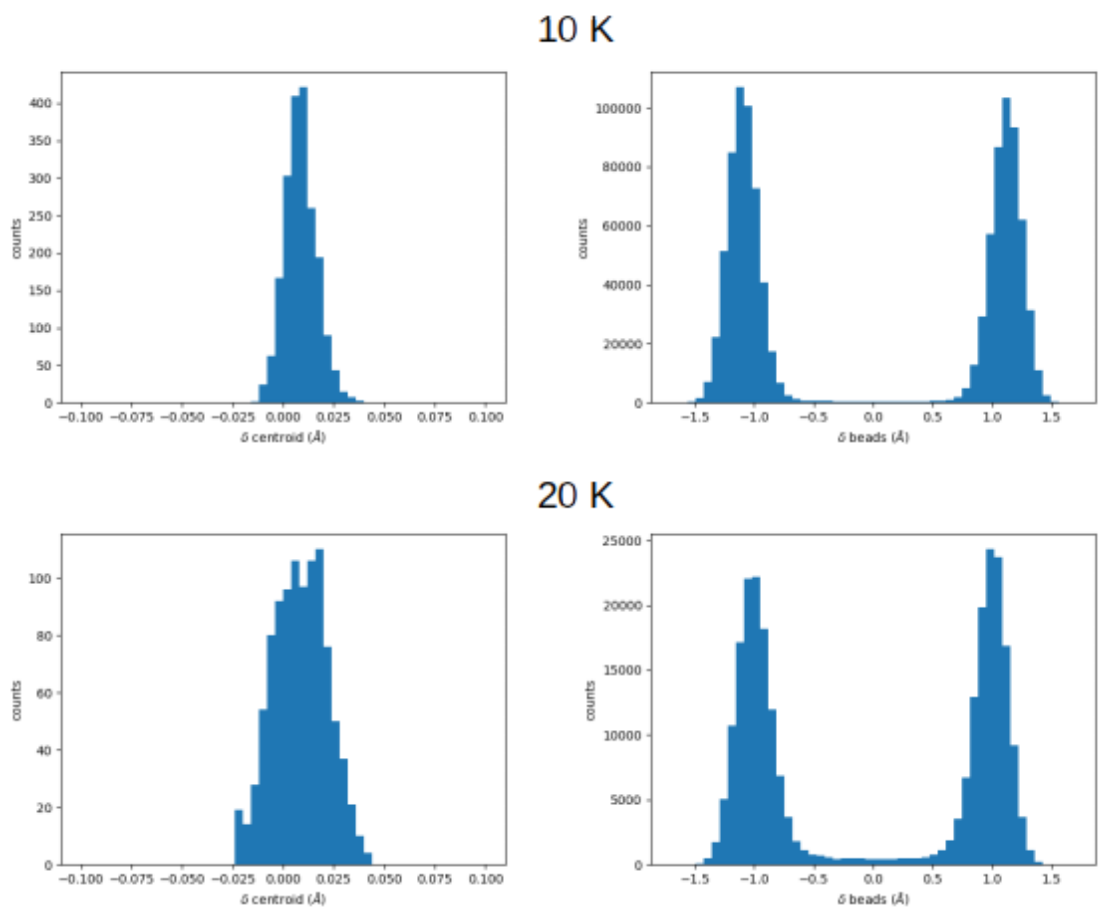

**Figure S16**: *Histogram of the distribution of the $\delta_2$ value for the centroid (left images), the $\delta_2$ for all the replicas (right images) at 10 K (top images) and 20 K (bottom images).*

When the centroid coordinate in the PIMD simulation has a value of δ=0 Å the state of the system is actually a combination of the two possible protonation states. In this way the system is able to maintain the π-conjugation in each of the protonated states.

Figure S17 displays histogram of distribution of δ-coordinates of all hydrogen atoms involved in the proton transfer mechanism obtained from the PIMD simulations at 10 K. The distribution shows strong correlation between positions of all hydrogen atoms, which is dictated by conservation of the π-conjugation. This demonstrates that the proton transfer is concerted.

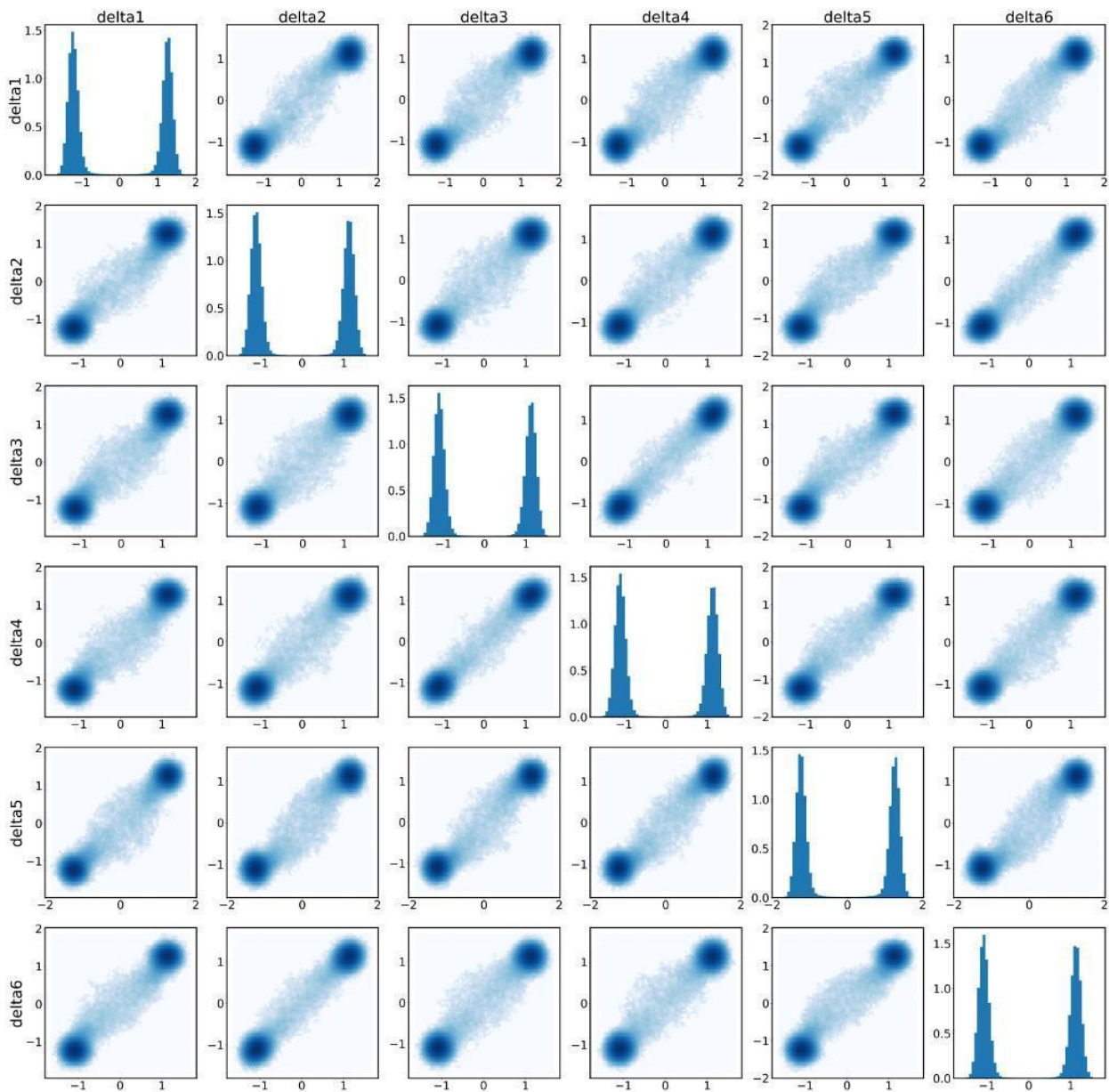

**Figure S17**: *Plot with the correlation between the δ-coordinates and the histogram with the distribution of each δ in the diagonal. $δ_1$ and $δ_4$ correspond to hydrogens in the edges and $δ_2$, $δ_3$, $δ_5$, $δ_6$ are hydrogens in the middle of the chain.*

Another way to see the correlation is to calculate the Pearson correlation coefficients that measure how far the reaction coordinates $\delta_i$ are from a linear correlation where 1.0 corresponds to perfect linear correlation and 0.0 means complete lack of correlation. In this matrix, representation we can see these values between each set of δ-coordinates in the PIMD simulations of the molecular trimer. The positions in the matrix correspond to the sub-index of the reaction coordinates δ shown in Figure S13.

|   | $\delta_1$ | $\delta_2$ | $\delta_3$ | $\delta_4$ | $\delta_5$ | $\delta_6$ |
|---|---|---|---|---|---|---|
| $\delta_1$ | 1.000 | 0.9826 | 0.9821 | 0.9823 | 0.9831 | 0.9833 |
| $\delta_2$ | 0.9826 | 1.000 | 0.9854 | 0.9840 | 0.9825 | 0.9866 |
| $\delta_3$ | 0.9821 | 0.9854 | 1.000 | 0.9865 | 0.9827 | 0.9854 |
| $\delta_4$ | 0.9823 | 0.9840 | 0.9865 | 1.000 | 0.9833 | 0.9847 |
| $\delta_5$ | 0.9831 | 0.9825 | 0.9827 | 0.9833 | 1.000 | 0.9832 |
| $\delta_6$ | 0.9833 | 0.9866 | 0.9854 | 0.9847 | 0.9832 | 1.000 |

**Table S1**: *Pearson correlation coefficients between distinct reaction coordinates $\delta_i$ obtained from the PIMD simulation.*

**Tight-Binding model on the molecular chain**

To construct the Tight-Binding model used in our analysis, we first considered an isolated molecule in the XY-plane and constructed its Hamiltonian with Fireball in the orthonormal Löwdin basis, using the BLYP functional for the calculation. We then selected the 10x10 matrix corresponding to the π-conjugated system (the Hamiltonian elements between the pairs of Löwdin π-orbitals associated to $p_z$ orbitals of carbon and nitrogen atoms). All the matrix elements between atoms not bonded covalently are 20 to 30 times smaller than those representing bonded atoms, so they are set to zero in our model. With $p_z$ orbitals labelled by the number of the atom (see Fig. S18), the resulting Hamiltonian $H_{tb,0}$ is (in the units of eV):

$$\begin{pmatrix} 6.02 & 0 & 3.21 & 0 & 0 & 0 & 0 & 0 & 0 & 0 \\ 0 & 3.48 & 0 & 3.65 & 0 & 0 & 0 & 0 & 0 & 0 \\ 3.21 & 0 & 2.77 & 2.75 & 3.47 & 0 & 0 & 0 & 0 & 0 \\ 0 & 3.65 & 2.75 & 2.28 & 0 & 3.06 & 0 & 0 & 0 & 0 \\ 0 & 0 & 3.47 & 0 & 2.32 & 0 & 3.06 & 0 & 0 & 0 \\ 0 & 0 & 0 & 3.06 & 0 & 2.33 & 0 & 3.47 & 0 & 0 \\ 0 & 0 & 0 & 0 & 3.06 & 0 & 2.28 & 2.75 & 3.65 & 0 \\ 0 & 0 & 0 & 0 & 0 & 3.47 & 2.75 & 2.77 & 0 & 3.21 \\ 0 & 0 & 0 & 0 & 0 & 0 & 3.65 & 0 & 3.46 & 0 \\ 0 & 0 & 0 & 0 & 0 & 0 & 0 & 3.21 & 0 & 6.02 \end{pmatrix}$$

**Table S2**: *Tight-binding Hamiltonian of the molecular unit obtained from Fireball DFT calculations. Labeling of sites with the molecular unit is represented in Figure S18.*

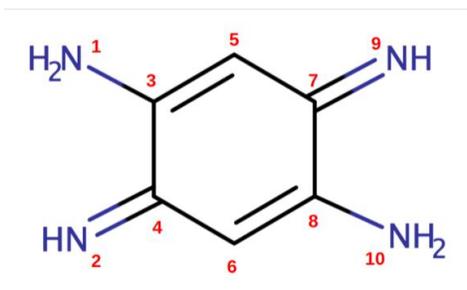

**Figure S18:** *Labeling used for the tight-binding Hamiltonian.*

To study a chain consisting of multiple units, we constructed a block matrix built up from replicas of $H_{tb,0}$ and added the hoppings associated to the hydrogen bonds connecting the neighboring units, $\tau_{hbond}$ (i.e. sites 9 and 1 of two adjacent blocks are coupled and similarly, sites 2 and 10). We assume that proton tunneling enhances this coupling compared to the *'frozen-*
43

*proton hydrogen bond*' case and also allows for a direct link between the terminal nitrogen pairs of the chain: a hopping $\tau_{edge}$ between the pairs of nitrogen atoms 1&2 in the first and 9&10 in the last molecule is added. We then proceed to explore our model over a wide range of values of $\tau_{edge}$ and $\tau_{hbond}$. The reason for this is that, at very low temperatures, the hydrogen transfer rate is heavily affected by nuclear quantum corrections, and thus cannot be correctly estimated by means of a DFT calculation relying on the Born-Oppenheimer approximation. We find that for situations in which $\tau_{edge} > \tau_{hbond}$, edge states appear in the gap as HOMO, HOMO-1, LUMO and LUMO+1 molecular levels, for values of $\tau_{edge}$ and $\tau_{hbond}$ which are small compared to the rest of the non-zero matrix elements (~0.5-1.5 eV), but still larger than the usual hopping for atoms bonded by hydrogen bonds (0.1 eV). This corroborates the importance of quantum nuclear effects at low temperature. The tunneling regime of the hydrogens leads to an effective interaction between the nitrogen atoms which is comparable in strength to a covalent bond.

Figure S19 summarizes the weights over the edges (defined as the two first and the last two molecules in the chain) of the HOMO and LUMO orbitals. The horizontal axis is the value of the hopping $\tau_{hbond}$, the vertical one is $\tau_{edge}$, and the color represents the projection of the wave function of the eigenstates on the sites that we identify as the edge regions. We also remark that our tight-binding model with $\tau_{edge}=0$ implies the existence of edge states related to the topologically-nontrivial states in the Su-Schrieffer-Heeger model,[45] but these never occur close to the Fermi level.

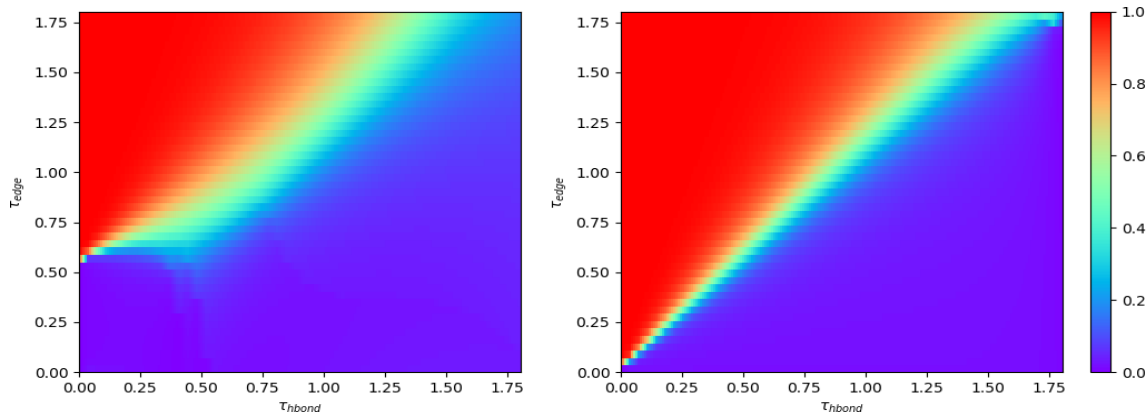

**Figure S19**: *Projection of the HOMO (a) and LUMO (b) to the edge region in our tight-binding model. Red color corresponds to the parameter region ($\tau_{edge}$, $\tau_{hbond}$) where the edge states occur.*

44. Cirera, B. *et al.* On-Surface Synthesis of Gold Porphyrin Derivatives via a Cascade of Chemical Interactions: Planarization, Self-Metalation, and Intermolecular Coupling. *Chem. Mater.* **31**, 3248–3256 (2019).

45. Su, W. P., Schrieffer, J. R. & Heeger, A. J. Soliton excitations in polyacetylene. *Phys. Rev. B* **22**, 2099–2111 (1980).